\documentclass[acmsmall, nonacm]{acmart}

\AtBeginDocument{
  \providecommand\BibTeX{{
    Bib\TeX}}}
  
\usepackage{verbatim}
\usepackage{tabularx}
\usepackage{makecell}
\usepackage{multirow}
\usepackage{multicol}
\usepackage{pgfplots}
\usepackage{listings}
\usepackage{xurl}
\usepackage{subcaption}
\usepackage{url}
\usepackage[normalem]{ulem}
\usepackage[linesnumbered,ruled,vlined]{algorithm2e}
\usepackage{footnote}
\usepackage{ragged2e}
\usepackage{enumitem}
\usepackage[most]{tcolorbox}
\usepackage[labelfont=bf]{caption}
\definecolor{darkgreen}{rgb}{0, 0.5, 0}
\definecolor{darkgreen1}{RGB}{37, 151, 19}
\definecolor{BurntOrange}{RGB}{203, 96, 21}
\definecolor{darkbrown}{rgb}{0.396, 0.263, 0.129}
\newcommand{\ignore}[1]{}

\def\BibTeX{{\rm B\kern-.05em{\sc i\kern-.025em b}\kern-.08em
    T\kern-.1667em\lower.7ex\hbox{E}\kern-.125emX}}
\def \tool{\textsc{Assertify}\xspace}
\def \toolNS{\textsc{Assertify}}
\lstdefinestyle{JavaStyle}{
    basicstyle=\footnotesize\ttfamily,
    keywordstyle=\color{blue},
    rulecolor=\color{black},
    commentstyle=\color{gray},
    stringstyle=\color{darkgreen1},
    keywordstyle=\color{blue},
    frame=single,
    breaklines=true,
    breakatwhitespace=true,
    showstringspaces=false,
    tabsize=2,
    lineskip=1.2ex,
    float
}

 \lstdefinestyle{InputGeneratorStyle}{
    language=java,
    basicstyle=\footnotesize\ttfamily,
    keywordstyle=\color{blue},
    rulecolor=\color{black},
    commentstyle=\color{gray},
    stringstyle=\color{darkgreen1},
    numbers=left,
    numberstyle=\tiny,
    numbersep=10pt,
    frame=none,
    breaklines=true,
    breakatwhitespace=true,
    showstringspaces=false,
    tabsize=2,
    lineskip=1.2ex,
    float
}

\acmDOI{}
\acmISBN{}

\begin{document}

\title{\toolNS: Utilizing Large Language Models to Generate Assertions for Production Code}

\author{Mohammad Jalili Torkamani}
\affiliation{
  \institution{University of Nebraska-Lincoln}
  \city{Lincoln}
  \state{Nebraska}
  \country{USA}
}
\email{mjalilitorkamani2@huskers.unl.edu}

\author{Abhinav Sharma}
\affiliation{
  \institution{Indraprastha Institute of Information Technology}
  \city{Delhi}
  \country{India}
}
\email{abhiji99@gmail.com}

\author{Nikita Mehrotra}
\affiliation{
  \institution{Microsoft}
  \city{New Delhi}
  \country{India}
}
\email{nmehrotra@microsoft.com}

\author{Rahul Purandare}
\affiliation{
  \institution{University of Nebraska-Lincoln}
  \city{Lincoln}
  \state{Nebraska}
  \country{USA}
}
\email{rahul@unl.edu}

\begin{abstract}
Production assertions are statements embedded in the code to help developers validate their assumptions about the code. They assist developers in debugging, provide valuable documentation, and enhance code comprehension. Current research in this area primarily focuses on assertion generation for unit tests using techniques, such as static analysis and deep learning. While these techniques have shown promise, they fall short when it comes to generating production assertions, which serve a different purpose. 

This preprint addresses the gap by introducing \tool, an automated end-to-end tool that leverages Large Language Models (LLMs) and prompt engineering with few-shot learning to generate production assertions. By creating context-rich prompts, the tool emulates the approach developers take when creating production assertions for their code. To evaluate our approach, we compiled a dataset of 2,810 methods by scraping 22 mature Java repositories from GitHub. Our experiments demonstrate the effectiveness of few-shot learning by producing assertions with an average ROUGE-L score of 0.526, indicating reasonably high structural similarity with the assertions written by developers. This research demonstrates the potential of LLMs in automating the generation of production assertions that resemble the original assertions.

\end{abstract}



\maketitle

\section{Introduction}
\label{sec:introduction}

\par Assertions are executable boolean expressions placed inside the program that must pass (return true) for all correct executions and fail (return false) for all incorrect executions\cite{10.1145/3368089.3409758}. They are recognized as a powerful tool to detect software errors in software testing and maintenance \cite{assertion_oriented_reference}. Production assertions are assertions that check whether certain conditions are met at runtime. They play an important role in documenting and debugging \cite{10.1145/143062.143098}. By embedding assertions within the code, developers enforce their assumptions about the code's behavior during runtime \cite{Autoassert}.

This makes it easier for developers and maintainers to grasp the underlying intent and logic behind the code, and use it for program debugging \cite{5948597} and maintenance \cite{10.1145/1321631.1321643}. 
Therefore, their importance in software engineering cannot be overstated, as they contribute to code quality \cite{Kudrjavets_paper} and maintainability \cite{10.1145/143062.143098}.

\par Despite the importance of these statements, generating effective production assertions remains a challenge. Significant research effort has been devoted to generating assertions for unit tests \cite{TOGA_2022, ATLAS_2020, AthenaTest, Autoassert, dinella}. These tools are specifically designed for unit test assertions. However, they are not readily applicable to generating production assertions. For instance, TOGA generates test oracles by analyzing the test prefix and the focal method (i.e., the method under test) as a context. A test prefix is a sequence of operations that sets up the initial state of the method under test before the actual test is performed, driving the method to a state where the test can then verify the expected behavior or outcome.

\begin{lstlisting}[style=JavaStyle,showstringspaces=false,tabsize=1,label=lst:toga_sample_input1,caption= ToGA Sample Input., captionpos=b, float=h, basicstyle=\scriptsize]
@kpublick@ @kstatick@ RealMatrix exp(@kfinalk@ RealMatrix matrix){
    @kreturnk@ @knewk@ Array2DRowRealMatrix(exp(matrix.getData()));
}
\end{lstlisting}
\vspace {-0.3cm}
\begin{lstlisting}[style=JavaStyle,showstringspaces=false,tabsize=1,label=lst:toga_sample_input2,caption= Generated Test Prefix., captionpos=b, float=h, basicstyle=\scriptsize]
@s// Test Prefixs@
@kdoublek@[][] data = {{1.0, 2.0},{3.0, 4.0}};
RealMatrix matrix = @knewk@ Array2DRowRealMatrix(data);
\end{lstlisting}
\vspace{-0.3cm}

\par To clarify further, consider the Java program in Listing \ref{lst:toga_sample_input1} \footnote{Extracted from the open-source repository https://github.com/finmath/finmath-lib}, which computes the exponential of a \texttt{RealMatrix} element-wise and returns a new \texttt{RealMatrix} as the result. As shown in Listing \ref{lst:toga_sample_input2}, TOGA generates a test prefix to set up the \texttt{RealMatrix} object with specific data to ensure that the matrix is in a specific state before the actual test is performed.

\begin{lstlisting}[style=JavaStyle,showstringspaces=false,tabsize=1,label=lst:toga_sample_input3,caption= Generated Test Case With Assertions., captionpos=b, float=h, basicstyle=\scriptsize]
@s// Test Case with Assertionss@
RealMatrix result = MatrixUtils.exp(matrix);
@kdoublek@[][] expected = {
    {Math.exp(1.0), Math.exp(2.0)},
    {Math.exp(3.0), Math.exp(4.0)}
};
assertArrayEquals(expected, result.getData());
\end{lstlisting}

After setting up the test prefix, TOGA generates the actual test case with assertions, as shown in Listing \ref{lst:toga_sample_input3}. However, for generating production assertions, we do not necessarily have or need a test prefix because the goal is to add assertions directly into the production code. This fundamental difference in approach makes TOGA and similar tools unsuitable for generating assertions for production code. Moreover, these tools do not consider the developer’s intended program behavior. They typically take unit tests as input and focus on the method under test, often relying on detecting undesirable output or exceptions (e.g., null dereferences, out-of-bound array accesses) as test oracles. This approach is insufficient for capturing the intended functionality of production code, which requires a deeper understanding of the context to generate meaningful assertions.

\par Another challenge in applying these techniques to generate production assertion is how they output the assertion statements. These tools generally do not specify the location for adding assertions, and within the unit tests, assert statements are often added toward the end. This is not the case with production assertions, which need to be integrated at appropriate points (line numbers) within the production code. Consequently, the generated test cases, while readable and accurate for unit tests, may not be integrated with the existing production codebase. An example of such difference is shown in Listing \ref{lst:toga_code_assertion_ful} where the production assertion checks for the nullability of the given input parameters and is located at the beginning of the method, the scenario that rarely occurs for unit test assertions.

\begin{lstlisting}[style=JavaStyle,showstringspaces=false,tabsize=1,label=lst:toga_code_assertion_ful,caption= Example of Production Assertion, captionpos=b, float=h, basicstyle=\scriptsize]
@kpublick@ @kstatick@ RealMatrix exp(@kfinalk@ RealMatrix matrix){
    @kassertk@ matrix != null;
    @kreturnk@ @knewk@ Array2DRowRealMatrix(exp(matrix.getData()));
}
\end{lstlisting}

\par To overcome these challenges, we propose \tool, that leverages Large Language Models (LLMs) \cite{10.1145/3641289} to generate assertions directly in production code. LLMs have demonstrated significant potential in assisting software engineers across various aspects of the development process. They excel at generating concise summaries of complex codebases and understanding programming languages and algorithms on par with human expertise \cite{exploring_llm_for_code_explanation, use_of_ai_driven_code_generation_models, comparing_code_explanations, experience_from_using_code_explanation, generating_diverse_code_explanations, automatic_generation_of_programming}. By harnessing LLMs' capabilities, particularly their understanding of code semantics and syntactic constructs, \tool generates and seamlessly integrates production assertions within the codebase, distinguishing itself from previous works that focus more on generating assertions for the unit tests.

\par \tool starts by extracting contextual information such as method name, signature, functional description, input/output details, etc. (detailed in Sections \ref{sec:approach} and \ref{sec:implementation}). Using this extracted context, \tool performs prompt engineering\cite{chen2024unleashingpotentialpromptengineering, white2023promptpatterncatalogenhance} and creates a specialized prompt designed to generate precise assertions for production code at specific locations. \tool demonstrates versatility in generating a wide range of assertions that capture the expected behavior, edge cases, and invariant conditions within the codebase. This approach ensures that the generated assertions are contextually relevant to the given method as input. We seek to address the following research questions:

\begin{enumerate} [leftmargin=*]
    \item \textbf{\textit{How syntactically accurate are the assertions generated by \tool?}}: 
    We explore the syntactical precision of the generated assertions using the SNE metric introduced in Section \ref{sec:experiment}.
    \item \textbf{\textit{How accurate are the static semantics of the assertions generated by \tool?}}: 
    To explore the static semantics of the generated assertions, we would check whether the resulting code after embedding the assertions is compilable or meaningful with respect to its static semantics and will calculate the SME metric introduced in Section \ref{sec:experiment}. Additionally, we will analyze the failed (not-compilable) cases and will classify them to identify the most frequent reasons for their failures.
    \item \textbf{How structurally similar are \tool-generated assertions compared to the original ones?}: 
    By `structure', we mean the organization of lexemes within the inserted assertion statements.
    We investigate and compare this structural similarity between the generated and original assertions using ROUGE scores.
 
\end{enumerate}

We make the following contributions:
\begin{enumerate} [leftmargin=*]
    \item We introduce \tool, an automated tool that leverages LLMs and prompt engineering to generate production assertions directly within codebases, addressing the limitations of existing unit test assertion generators.
    \item We develop a methodology for integrating context-aware assertion generation using few-shot learning, enhancing the relevance and accuracy of the assertions produced.
    \item We create and share a dataset of 2,810 methods from 22 mature Java repositories, including 983 methods with 1,548 developer-written assertions for evaluation, to facilitate future research in this area.
\end{enumerate}

\par This preprint paper is organized as follows: Section \ref{sec:motivation} discusses our motivation behind conducting this research. Section \ref{sec:background} briefly describes the chain-of-thought and few-shot learning topics that are related to our research. Section \ref{sec:approach} discusses the approach we followed, while Section \ref{sec:implementation} describes how we built the \tool to implement the approach. Section \ref{sec:experiment} explains how we collected our dataset and the experiments we designed to resolve each research question. In Section \ref{sec:results} we present a thorough discussion and interpret the results of our conducted experiments. Potential threats to the validity of our findings are discussed in Section \ref{sec:threatstovalidity}. In addition, Section \ref{sec:relatedwork} discusses some of the conducted related works in the assertion and code generation area, categorized by the approach they follow. Finally, the preprint concludes in Section \ref{sec:conclusion}, summarizing key findings, their significance, and future works. Section \ref{sec:data_availability} provides a link to the research artifacts.
\section{Motivation}
\label{sec:motivation}

Production assertions assist developers in program comprehension and debugging. They provide insights into how statements operate and specify the intended behavior of methods. Additionally, they allow developers to recognize the violated assumptions when a failure is reported in a production environment \cite{341844}. Despite the advantages of these programming statements, there is still a significant gap in their usage across open-source repositories. A closer look at GitHub repositories using SourceGraph \cite{sourceGraph}\footnote{SourceGraph is a web-based code search engine enabling developers to search for code snippets, navigate codebases, and understand code dependencies. Additionally, it supports the extraction of GitHub repositories using criteria like code regex, file path, and programming language.} reveals that many open-source repositories do not incorporate assertions within their methods.

\par To substantiate this claim, we used an assertion-finding regex pattern \verb|^[^(//)*]\sassert\s+\w*| and selected the Java programming language and the path \textit{src/main/java}. Our analysis revealed over 946 repositories had production assertion occurrences when applying the production assertion regex, whereas, without this pattern, that was 38,006 repositories. These findings suggest that only about 2.48\% of the repositories had production assertions. This low rate is not surprising since developers often avoid using assertions to save development time, potentially compromising the reliability and maintainability of their code. However, it is imperative to approach these statistics with caution. Typical large-scale Java projects involve numerous methods, sometimes numbering in the hundreds or thousands, making the manual generation of production assertions a burden and, in most cases, impractical. Moreover, it also requires domain-specific knowledge that not all developers possess.

\par These challenges led us to explore a solution for generating production assertions. We designed a novel approach that generates assertions for the given input methods, leveraging the capabilities of LLMs for generating assertions. 
\begin{lstlisting}[style=JavaStyle,showstringspaces=false,tabsize=2,label=lst:mot_original_code,caption= Original Code.,captionpos=b, float=h, basicstyle=\scriptsize]
@Override
@kpublick@ void write(ShortBuffer dest, ObjectToOffsetMapping mapping) {
    writeFirst(AA, dest);
    @sassert (getOffset() + BBBBBBBB) % 2 == 0;s@
    write32BitValue(BBBBBBBB, dest);
}


\end{lstlisting}

    \par To illustrate our approach further, consider the code snippet in Listing \ref{lst:mot_original_code} extracted from a large-scale GitHub repository\footnote{\url{https://github.com/virjarRatel/ratel-core}}. In the \texttt{write} method, the original assertion (highlighted in green) verifies whether the sum of the variable \texttt{BBBBBBBB} and the output from the \texttt{getOffset} method is even.

\par To replicate this process and generate reasonable assertions, the first approach could be using LLMs without providing much context. Considering a naive prompt \textit{A} (provided in Step 1 in Figure \ref{fig:sample_prompt_creation}), the resulting code with a new set of assertions is shown in Listing \ref{lst:mot_naive_code_a}\footnote{The provided inference has been extracted from our experiments' results.}:

\begin{lstlisting}[style=JavaStyle,showstringspaces=false,tabsize=2,label=lst:mot_naive_code_a,caption= Assertify Inference (Prompt A / GPT-4).,captionpos=b, float=h, basicstyle=\scriptsize]
@Override
@eassert dest != null;e@
@eassert mapping != null;e@
@kpublick@ void write(ShortBuffer dest, ObjectToOffsetMapping mapping) {
    writeFirst(AA, dest);
    write32BitValue(BBBBBBBB, dest);
}
\end{lstlisting}

\par At first glance, it is evident that the resulting code after inserting assertions (highlighted in red) has some serious syntactic issues. Due to the embedding of two assertions in a place that violates Java grammar, the resulting code is not parsable. Moreover, even after fixing this issue, the code must be compiled to verify if the resulting code has any static semantic errors (in this case, it is compilable).

\par To improve the initial approach further, we refine our method and provide the LLM with more context using a new prompt \textit{B} (provided in Step 2 in Figure \ref{fig:sample_prompt_creation}), which included a code summary. After making an inference for the same method, the resulting code with added assertions is shown in Listing \ref{lst:mot_naive_code_b}.

\begin{lstlisting}[style=JavaStyle,showstringspaces=false,tabsize=2,label=lst:mot_naive_code_b, caption=Assertify Inference (Prompt B / GPT-4).,captionpos=b, basicstyle=\scriptsize, float=h]
@Override
@kpublick@ void write(ShortBuffer dest, ObjectToOffsetMapping mapping) {
    @sassert dest != null;s@
    @sassert mapping != null;s@
    writeFirst(AA, dest);
    @eassert AA != null;e@
    write32BitValue(BBBBBBBB, dest);
    @eassert BBBBBBBB != null;e@
}
\end{lstlisting}

\par As depicted in Listing \ref{lst:mot_naive_code_b}, the syntactic problem appears to be fixed using the improved prompt. However, there are still some errors that would be revealed only after compiling the repository. These issues are of static semantics highlighted in red, where two integer variables are mistakenly treated as objects and compared against null. Therefore, although the syntactic issue has been resolved by adding context, there remain two (static-) semantic errors that have yet to be solved.

\begin{lstlisting}[style=JavaStyle,showstringspaces=false,tabsize=2,label=lst:mot_rich_code_e,caption= Assertify Inference (Prompt E / GPT-4).,captionpos=b, float=h, basicstyle=\scriptsize]
@Override
@kpublick@ void write(ShortBuffer dest, ObjectToOffsetMapping mapping) {
    @sassert dest != null && mapping != null;s@
    writeFirst(AA, dest);
    write32BitValue(BBBBBBBB, dest);
}
\end{lstlisting}

\par The ultimate approach  (Step 4 in Figure \ref{fig:sample_prompt_creation}) generally involves continuing prompt engineering \cite{reynolds2021promptprogramminglargelanguage,kang2023largelanguagemodelsfewshot} further by providing more context, including the method's input(s) and output, the code summary of the invoked dependencies, and more importantly, providing structurally similar methods to the LLM using few-shot learning. The LLM would then be furnished with a rich and well-crafted input prompt, generating more meaningful assertions, free of syntactic and static semantic errors, as shown in Listing \ref{lst:mot_rich_code_e}.

\par In summary, our motivation lies in automating the process of generating production assertions by harnessing prompt engineering techniques in LLMs to imitate the process developers follow when generating assertions. This relieves developers of the manual burden of writing production assertions, saving time and increasing the reliability of codebases.
\section{Background}
\label{sec:background}

\subsection{Chain-of-thought prompting}

Chain of Thought (CoT)\cite{wei2023chainofthought} is an approach in Natural Language Processing (NLP) that enhances the capabilities of LLMs to solve complex problems and perform reasoning tasks. This method involves structuring the problem-solving process into intermediate steps, allowing the model to ``think aloud" as it navigates through a sequence of logical deductions before arriving at a final answer. By explicitly articulating these steps, CoT empowers LLMs to tackle tasks that require more profound understanding and multi-step reasoning, ranging from arithmetic problems to complex comprehension questions. This methodology improves the transparency of how LLMs reach their conclusions, significantly enhances their ability to handle various tasks, and orients them toward imperative solving procedures \cite{ye2023satlm}. It marked a pivotal shift in the development of AI, opening new pathways for research and application in various domains, including software engineering.

\subsection{Few-shot learning}
Few-shot learning \cite{10.1145/3386252,parnami2022learningexamplessummaryapproaches} is a technique in machine learning, specifically within the context of LLMs. This approach leverages a minimal set of examples to guide LLMs in executing specific tasks, demonstrating how to perform a task with only a few data samples. In contrast to conventional machine learning techniques that demand large datasets for training, few-shot learning leverages LLMs' pre-existing knowledge and flexibility, allowing them to adapt their learned patterns and contextual understanding to novel challenges with minimal explicit guidance. This approach not only simplifies the training process but also highlights the LLMs' capacity for quick learning and their adeptness in handling a wide range of tasks with new information. The implications of few-shot learning for AI development are profound, offering a path to augmenting model capabilities and broadening their use in niche areas without necessitating extensive data collection or training efforts.
\section{Approach}
\label{sec:approach}

\begin{figure*}
    \centering
    \includegraphics[width=\textwidth]{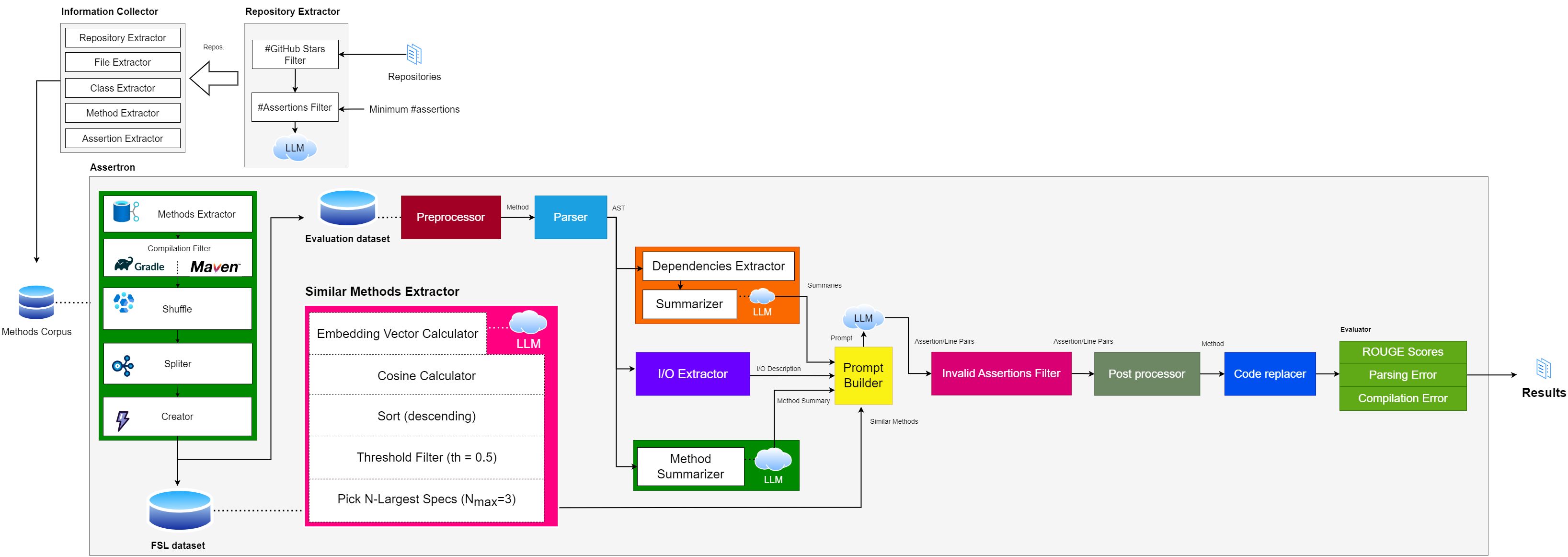}
    \caption{\textbf{\tool Workflow.}}
    \label{fig:pipeline_diagram}
\end{figure*}

\par Our approach draws inspiration from developers' actions to generate assertions. It starts with extracting information blocks from the input and benefits from prompt engineering to use them as context. By intricately weaving contextual information into the input prompts, the LLM is equipped with the essential and nuanced context. This context is required to generate the intended assertions. Compared to the chain-of-thought \cite{wei2023chainofthought} methodology, which provides an inferential path in the expected response during few-shot prompting, our approach enriches the input prompt and resembles how developers craft production assertions.

In this section, we present a detailed perspective on our methodology, adopting the previous code fragment (Listing \ref{lst:mot_original_code}) to explain the entire process. 

\par Figure \ref{fig:pipeline_diagram} depicts \tool's overview. First, input records, including candidate methods\footnote{The method for which we want to generate production assertions.} and their associated metadata, are extracted from the dataset and trimmed by \textit{preprocessor} to remove all assertions and comments that could distract the LLM from generating unbiased responses (Listing \ref{lst:mot_pruned_code}). These metadata elements will be thoroughly described in Section \ref{sec:implementation}, where we talk about how they are extracted from our input dataset. The \textit{preprocessor} component also duplicates the repository which will be used later by the \textit{Evaluator} component.

\begin{lstlisting}[style=JavaStyle,showstringspaces=false,tabsize=2,label=lst:mot_pruned_code, caption=Pruned Method.,captionpos=b, float=h, basicstyle=\scriptsize]
@Override
@kpublick@ void write(ShortBuffer dest, ObjectToOffsetMapping mapping) {
    writeFirst(AA, dest);
    @kassertk@ (getOffset() + BBBBBBBB) % 2 == 0;
    write32BitValue(BBBBBBBB, dest);
}
\end{lstlisting}
 
\begin{figure*}
    \centering
    \includegraphics[width=\textwidth]{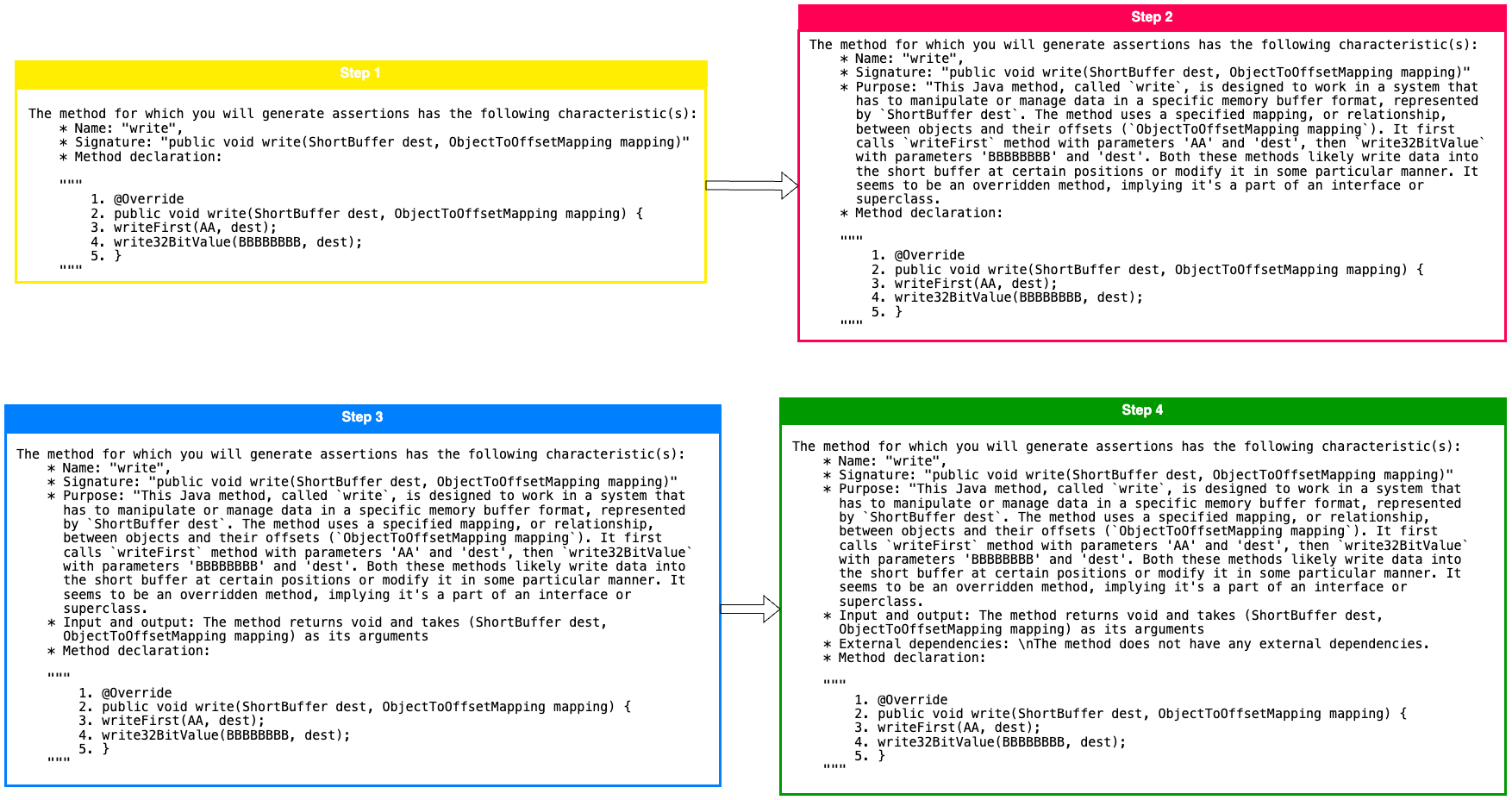}
    \caption{\textbf{Example of Prompt Creation.}}
    \label{fig:sample_prompt_creation}
\end{figure*}

Next, the trimmed method (called the pruned method) undergoes a prompt engineering process, exemplified in Figure \ref{fig:sample_prompt_creation}. This process is crucial, as it directly impacts the quality of the generated assertions, leading to higher accuracy if they are well-established. We outline the steps our approach takes to generate production assertions, along with the reason behind each step.

The process begins by extracting the method's name and signature using metadata and embedding the result in the final prompt (Step 1). 

Given that developers often turn their attention to semantics, aiming to precisely align generated assertions, an appropriate prompt is created to guide the LLM in generating code summaries. The effectiveness, correctness, and accuracy of these generated code summaries have been corroborated by recent studies, e.g., \cite{experience_from_using_code_explanation}, illustrating the benefits of employing LLMs for generating method explanations. This code summary ensures that the information integrated into the prompt helps the LLM understand the method's intended behavior (Step 2).

Going a step further, \tool also incorporates information about the method's inputs and outputs into the prompt. This additional data enables the LLM to consider the expected behavior of the method, just as developers do manually when constructing assertions (Step 3).

When developers manually write assertions, they understand the method's external dependencies, offering semantic insights and enriching the assertion-writing process. To implement this, \tool extracts the source code of internally defined methods invoked within the candidate method, integrates them into an input prompt, and feeds it into the LLM to obtain code summaries expressing their objectives. The generated code summaries, along with the method's name, signature, and inputs/outputs, are integrated into the prompt. This approach ensures a comprehensive understanding of the context for the method for which we generate assertions, as well as its internally invoked methods (Step 4).

After step 4, \tool integrates the extracted elements as a single prompt using the \textit{Builder} component, which encapsulates essential information for the LLM. Since developers often derive inspiration from previously seen methods with assertions, \tool employs a few-shot learning technique. Using the \textit{Similar Method Extractor} component, it selects the top three methods most similar to the candidate method and incorporates them into the existing prompt, forming the final prompt to be sent to the LLM. (We will explain it in detail in Section \ref{sec:implementation})

\par After the LLM generates a set of assertions along with their associated line numbers, this data is passed to the \textit{postprocessor} component. The \textit{postprocessor} filters out assertions where the predicted line numbers are incorrect. The correctness of line numbers is verified by ensuring that they fall within the method’s boundaries—specifically between the method's starting and ending curly braces.

\par The remaining assertions are then inserted into the candidate method at their corresponding predicted line numbers. The resulting method (having inserted assertions) replaces the original in the codebase using metadata that helps locate the exact method in the dataset. This process ensures the LLM-generated assertions are incorporated without concerns about any overloaded Java methods.

\par Next, the new method, along with the assertions and their corresponding line numbers, is sent to the \textit{evaluator} component for analysis. The \textit{evaluator} performs both syntactic and semantic evaluations. To carry out these assessments, the \textit{evaluator} compares the new assertions against the other duplicate of the repository, which the \textit{preprocessor} created earlier. Various metrics are measured before and after the code replacement, as detailed in Section \ref{sec:experiment}. These metrics are crucial for evaluating the responses. Additionally, the \textit{evaluator} component tracks inference time and costs, offering insights into the efficiency of each model in generating production assertions. It also provides valuable information regarding inference performance and resource expenditure. During evaluation, the entire method is parsed to detect any syntactic errors. Next, static semantic errors are checked by compiling the entire repository having the method replaced with the older version, to detect any static semantic issues introduced by the assertions. The linguistic similarity between the original assertions (used as the reference document) and the generated assertions (used as the candidate document) is measured using ROUGE-L scores. Finally, all evaluation results and information from previous components are saved in a JSON file for later analysis.
\section{Implementation}
\label{sec:implementation}

\tool employs a hybrid approach, integrating Java for core functionalities with Python to perform prompt tokenization, interact with LLM, and compute average ROUGE scores. It uses \textit{JavaParser} library to parse Java code and utilizes GPT-3.5-turbo-16k (hereafter referred to as GPT-3), GPT-4 model, and GPT-4o model as LLM, aiming to explore their capabilities in generating production assertions. The implementation is divided into three phases:
\begin{itemize}[leftmargin=*]
    \item \textbf{Repositories Extraction:} Utilizing SourceGraph, we collect large-scale open-source targeted Java repositories containing production assertions, followed by filtering them using GitHub APIs to eliminate repositories failing to meet predefined criteria (i.e., number of stars and number of production assertions), and cloning the selected repositories for subsequent analysis. We will talk about the statistics in Section \ref{sec:experiment} where we detail the data collection operations.   
    \item \textbf{Metadata Creation:} We generate a rich JSON metadata file, named \textit{methods corpus}, cataloging file paths, classes, methods, comments, and assertions, facilitating data retrieval by circumventing the need for full repository scans as any information is needed. This leads to reducing computational overhead and saves time while generating assertions.
    \item \textbf{\tool Pipeline:} Our approach, designed as a pipeline with sequential modular components, generates pairs of production assertions and line numbers (at which they should be inserted) and evaluates them using syntactic and static semantic accuracy criteria. All the steps within the pipeline are automatically performed, eliminating the necessity of any manual intervention.
\end{itemize}

For few-shot learning, the \textit{Similar Method Extractor} component selects the top three similar methods. This choice balances the need for sufficient contextual examples for the LLM with minimizing token consumption to stay within the model's input limits. To compute similarity during prompt engineering, \tool vectorizes both the candidate method and each unseen method\footnote{An unseen method refers to a method in the few-shot learning dataset (FSL) that is never used as a candidate method.}, padded with zeros to make them equal in size. Subsequently, it applies the cosine similarity formula, sorts the results in descending order, and sets 0.5 as the minimum similarity threshold: 
\[
\text{{cosine similarity}}(\mathbf{A}, \mathbf{B}) = \frac{\mathbf{A} \cdot \mathbf{B}}{\|\mathbf{A}\| \times \|\mathbf{B}\|}
\]

This threshold strikes a balance between precision and recall, ensuring a diverse set of methods covering a wide range of similarity levels, from lower similarities serving as false positives to higher similarities. Regarding the LLM inference, the tool can benefit from either Application Programming Interfaces (APIs) (through sending web requests) or deployed offline LLMs. All three models use the default configuration when making any inference. Correspondingly, the LLM is provided with a set of \textit{system} instructions \footnote{The instructions are available in our released artifacts} to guide the LLM toward generating expected assertions, while the constructed prompt is given as \textit{user} field. To provide similar examples, pairs of \textit{user}-\textit{assistant} fields are presented to the LLM to facilitate few-shot learning, allowing the fourth pair to rely solely on the \textit{user} for inference based on the given examples. In addition, due to the internal constraints of models in terms of number of tokens, the final prompt is tokenized beforehand to check if the number of tokens adheres to the LLM's limitations. Otherwise, few-shot learning samples will be decreased until the prompt size meets the LLM's associated input range.

Regarding the insertion of assertions, after making an inference, the \tool iterates over each assertion-line number pair, inserting each assertion at its corresponding predicted line number. Note that line numbers start from 1 and are based on the pruned method, with all line numbers measured against the line numbers of this method. To check whether the resulting method has any static semantic errors, the \tool compiles the enclosing repository using the appropriate Java version, considering the exit code as the compilation result.

Therefore, this approach includes dataset creation, prompt generation, and response evaluation, contributing toward generating assertions. \tool also aligns the repository of each dataset record (i.e., method) with a corresponding Java version during compilation. This intentional choice minimizes compilation errors by acknowledging the variability in project requirements and dependencies. Furthermore, utilizing APIs for making inferences enhances the accessibility and scalability of the tool without struggling with implementations of LLMs or heavy computational resources for inference. 
\section{Experiment Design}
\label{sec:experiment}

In this section, we explain the experiments performed to answer each research question mentioned in Section \ref{sec:introduction}. We begin by describing the dataset collection process and providing relevant statistics. Next, we discuss the experiment we conducted, and the metrics used to evaluate our results. Finally, we describe how we use each metric to address each research question.

\subsection{Dataset Collection}
\par Our experiments began by collecting 3,819 unique Java repositories from GitHub using SourceGraph, filtered by \textit{src/main/java} file paths. We applied the regex pattern `\verb|^[^(//)*]\sassert\s+\w*|' to precisely identify repositories containing at least one production assertion. This approach allowed us to focus on Java projects with production assertions, reducing the need to analyze all repositories. The resultant report file filtered out less mature or less popular projects by requiring at least five hundred stars, narrowing the repository pool to 1,098. We further refined the selection to include only repositories with at least 50 production assertions, resulting in 22 mature and popular repositories with a sufficient number of production assertions. We established these minimum thresholds after analyzing their distribution across all selected repositories, ensuring the inclusion of relatively mature repositories with extensive use of production assertions. Table \ref{tab:dataset-stats-table} provides detailed statistics on our datasets, including the number of repositories, code elements (classes, methods, and variables), and input datasets.

\begin{table}[t]
\small
\centering
\caption{\textbf{Dataset Statistics.}}
\renewcommand{\arraystretch}{1.15}
\setlength{\tabcolsep}{8pt} 
\begin{tabular}{|c|c|c|}
\hline
\textbf{Metric} & \multicolumn{2}{c|}{\textbf{Value (\#)}} \\ \hline
\multirow{2}{*}{\textbf{Repositories}} & \textbf{Total} & \textbf{Filtered} \\ \cline{2-3}
   & 3,819 & 22 \\ \hline
\multirow{2}{*}{\textbf{Files}} &\textbf{ Total} & \textbf{With Assertions} \\ \cline{2-3}
   & 131,445 & 1,401 \\ \hline
\multirow{2}{*}{\textbf{Classes}} & \textbf{Total} & \textbf{With Assertions} \\ \cline{2-3}
   & 137,848 & 2,674 \\ \hline
\multirow{2}{*}{\textbf{Methods}} & \textbf{Total} & \textbf{With Assertions} \\ \cline{2-3}
   & 1,076,084 & 2,810 \\ \hline
\multirow{2}{*}{\textbf{Variables}} & \textbf{Total} & \textbf{Used by Assertions} \\ \cline{2-3}
   & 909,412 & 5,742 \\ \hline
\textbf{Assertions} & \multicolumn{2}{c|}{4,412} \\ \hline
\multirow{2}{*}{\textbf{Dataset Split}} & \textbf{Evaluation Set} & \textbf{FSL} \\ \cline{2-3}
 & 983 & 1,827 \\ \hline
\end{tabular}
\label{tab:dataset-stats-table}
\end{table}

\par After extracting repositories, we created and stored the method metadata in structured JSON format. From the curated collection of over 131,445 Java non-test source code files, we identified 1,401 methods containing assertions within a larger corpus totaling 2,810 methods. We counted the Java files of each repository while iterating over non-test directories, then partitioned this collection into two distinct datasets for few-shot learning and evaluation. \tool uses the first dataset to extract similar methods when prompt engineering and the latter is used to evaluate the accuracy. Using the \textit{JavaParser} library to parse the source codes, we selected 983 candidate methods for our evaluation set, focusing on methods containing at least one assertion. For the few-shot learning dataset, 1,827 methods were chosen from the remaining pool of methods with assertions. These methods were crucial for training \tool to identify analogous methods. Table \ref{tab:dataset-distribution-table} shows the distribution of assertions per file, class, and method, as well as the distribution of classes, methods, and variables used within the repositories' original assertions.

\begin{table}[h]
\centering
\caption{\textbf{Assertion Distribution.}}
\renewcommand{\arraystretch}{1.15}
\setlength{\tabcolsep}{5pt} 
\begin{tabular}{|c|c|c|c|}
\hline
\textbf{Metric} & \multicolumn{3}{c|}{\textbf{Distribution (\%)}} \\ \hline
\multirow{2}{*}{\textbf{Location Distribution}} & \textbf{Per File} & \textbf{Per Class} & \textbf{Per Method}\\ \cline{2-4}
 & 1.833 & 1.939 & 0.617 \\ \hline
\multirow{2}{*}{\textbf{Target Distribution}} & \textbf{Per Class} & \textbf{Per Method} & \textbf{Per Var.}\\ \cline{2-4}
 & 2.148 & 0.027 & 0.631 \\ \hline
\end{tabular}
\label{tab:dataset-distribution-table}
\end{table}

\subsection{Experimental Methodology}
We detail the methodology used to evaluate the quality of assertions generated by \tool. We ran \tool with three models (GPT-3, GPT-4, and GPT-4o) and five prompt templates, resulting in 15 variations of experiments. Each variation included different contextual information to guide the model's inference, allowing us to determine which was more effective.

\begin{enumerate} [leftmargin=*]
    \item \textbf{Prompt \textit{A}}: A naive prompt includes only the method name, signature, and body. We will use this prompt as a base and will augment it with additional context.
    \item \textbf{Prompt \textit{B}}: Extends \textit{A} by adding a code summary.
    \item \textbf{Prompt \textit{C}}: Extends \textit{B} by adding input-output descriptions.
    \item \textbf{Prompt \textit{D}}: Extends \textit{C} by adding summaries of invoked methods' code.
    \item \textbf{Prompt \textit{E}}: The most comprehensive prompt, extending \textit{D} by adding similar methods using few-shot learning.
\end{enumerate}

We define the following evaluation metrics, to answer each research question:

\begin{itemize} [leftmargin=*]
\item \textbf{Syntactic Error (SNE) Rate:} This metric measures the proportion of methods with syntactic errors (after inserting inferred assertions at their associated line numbers) that fail to parse successfully, over the total number of the evaluation set. Such errors include any syntactic errors that occur after adding assertions, including missing semicolons, malformed assertions (violating Java grammar), and invalid assertion locations that result in failure in parsing the method. The SNE rate evaluates \tool in generating assertions that do not comply with Java grammar. 

\item \textbf{Semantic Error (SME) Rate:} This metric evaluates the proportion of methods with static semantic errors, over the total number of the evaluation set. Unlike the SNE rate, SME covers compilation issues beyond parsing, such as scope rules, type checking, naming conflicts, and poorly structured control flow statements. As the method statements remain unchanged after adding assertions, the static semantics should be preserved. This metric does not measure errors that may arise after execution (i.e., no dynamic analysis or execution of the methods is involved within this metric.).

\item \textbf{ROUGE Score:}
We use the ROUGE score \cite{lin2004rouge} as our final evaluation metric. We report the ROUGE-L score which reflects the linguistic structural similarity for the longest common subsequence rather than correctness or functionality. We chose this metric over ROUGE-N scores because the tokens can be combined at various levels, making ROUGE-N inappropriate for assessing structural similarity at the expression level rather than the token level. In our context, precision reflects the syntactic relevance of generated assertions, while recall indicates their syntactic completeness compared to original assertions. Higher $\text{F}_1$-scores imply greater syntactic similarity and closer linguistic resemblance between generated and original assertions.

\[
\text{Precision (P)} = \frac{\text{overlapping n-grams}}{\text{total generated n-grams}}
\]
\[
\text{Recall (R)} = \frac{\text{overlapping n-grams}}{\text{total reference n-grams}}
\]
\[
\text{F}_1-Score~(F) = \frac{2\text{PR}}{\text{P+R}}
\]

\end{itemize}

\subsubsection{\textbf{\textit{RQ1: How syntactically accurate are the assertions generated by \tool?}}}

This research question evaluates the syntactical correctness of generated assertions and their insertion line numbers for a given code fragment\footnote{We use method-level granularity throughout this research.} from the dataset. We use the SNE rate to assess the syntactic accuracy of the method after assertions are inserted. Here, syntactic accuracy refers to the parsability of the code after placing predicted assertions in predicted line numbers. Thus, any syntactically incorrect assertion or mispredicted insertion line that makes the resulting code unparsable will decrease syntactic accuracy and increase the SNE rate. 

\subsubsection{\textbf{\textit{RQ2: How accurate are the static semantics of the assertions generated by \tool?}}}

This research question evaluates \tool's accuracy in generating assertions with correct static semantics. To answer this, we use the SME rate, which involves compiling the entire repository after replacing each method with its updated version. Additionally, we analyze the main reasons behind these static semantic errors after embedding the generated production assertions. We classify these failures to identify the most frequent errors encountered by \tool in generating assertions for a given method.

\subsubsection{\textbf{\textit{RQ3: How structurally similar are \tool-generated assertions compared to the original ones?}}}

To answer this research question, we compare the generated assertions to the original ones for each method. Since the mapping between assertions is not often strictly one-to-one, the tool may generate additional assertions, omit existing ones, or redistribute them, leading to variations. Therefore, we use average ROUGE-L scores to measure structural similarity, which, in turn, is a measure of lexical sequence alignment. Unlike ROUGE-1 and ROUGE-2 scores, which compute unigrams and bigrams similarities, the ROUGE-L score measures the similarity for the Longest Common Subsequence (LCS), providing a better understanding of how similar two sets of assertion expressions are. Higher average ROUGE-L scores indicate greater structural similarity between the sets of assertions. Even though this metric does not directly measure the dynamic semantic correctness of the generated assertions, it determines the extent to which the \tool generated set of assertions is similar to the set of original assertions written by developers. We also measure the weakness, equivalence, and strength of the generated assertions to gain further insight into this issue. To do so, we determine whether the set of original assertions is a subset, equal to, or a superset of the set of \tool generated assertions.

Additionally, we compare the performance of the \tool when using the GPT-3, GPT-4, and GPT-4o models in terms of time and cost, providing developers with insights regarding choosing the appropriate LLM for generating assertions.

\section{Results}
\label{sec:results}

\begin{figure*}[t] 
    \centering
    \begin{minipage}[b]{0.49\textwidth}
        \centering
        \includegraphics[width=\textwidth]{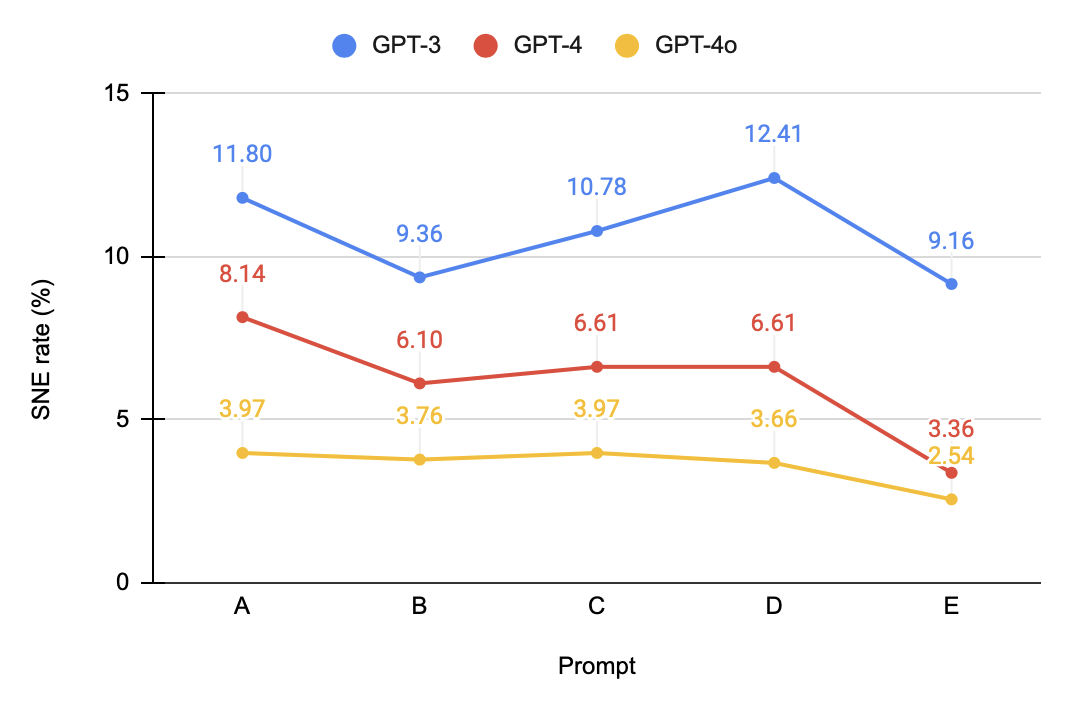}
        \caption{SNE Rate.}
        \label{fig:sne_rate}
    \end{minipage}
    \hfill
    \begin{minipage}[b]{0.49\textwidth}
        \centering
        \includegraphics[width=\textwidth]{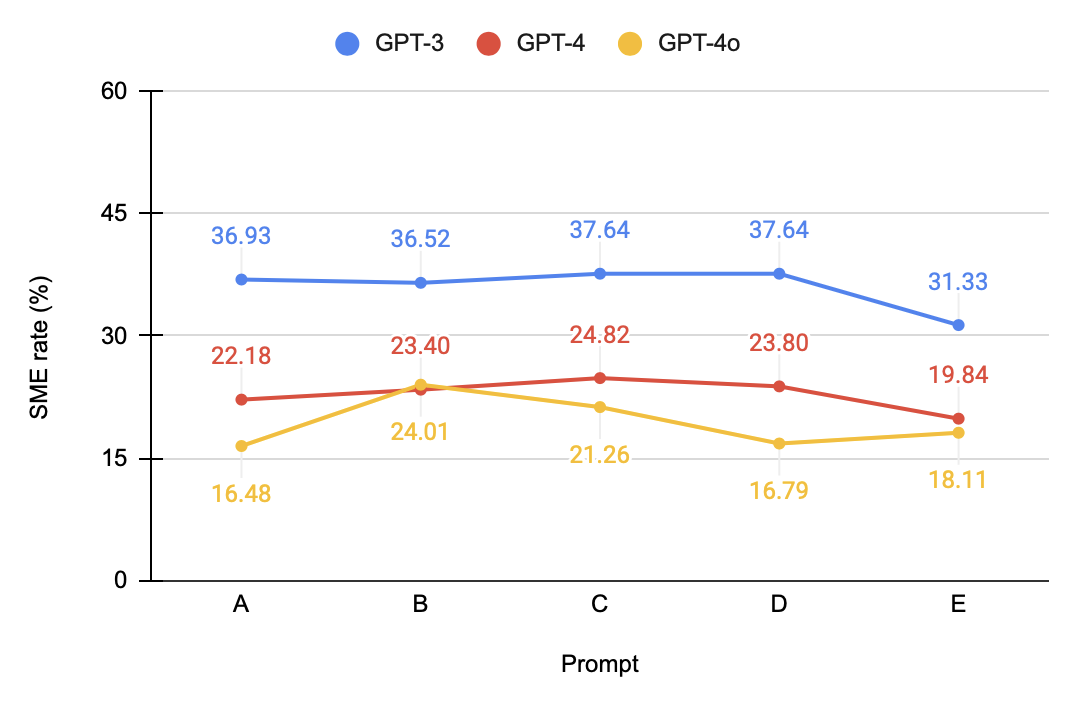}
        \caption{SME Rate.}
        \label{fig:sme_rate}
    \end{minipage}

    \vspace{0.3cm} 

    \begin{minipage}[b]{0.49\textwidth}
        \centering
        \includegraphics[width=\textwidth]{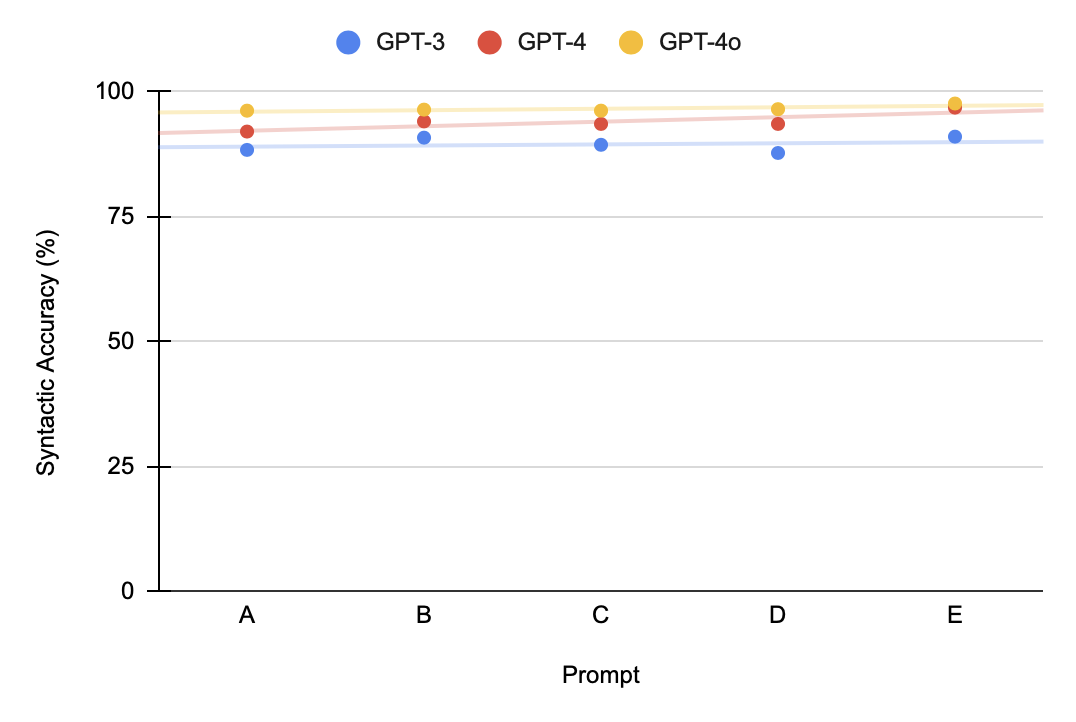}
        \caption{Syntactic Accuracy.}
        \label{fig:syntactic_accuracy}
    \end{minipage}
    \hfill
    \begin{minipage}[b]{0.49\textwidth}
        \centering
        \includegraphics[width=\textwidth]{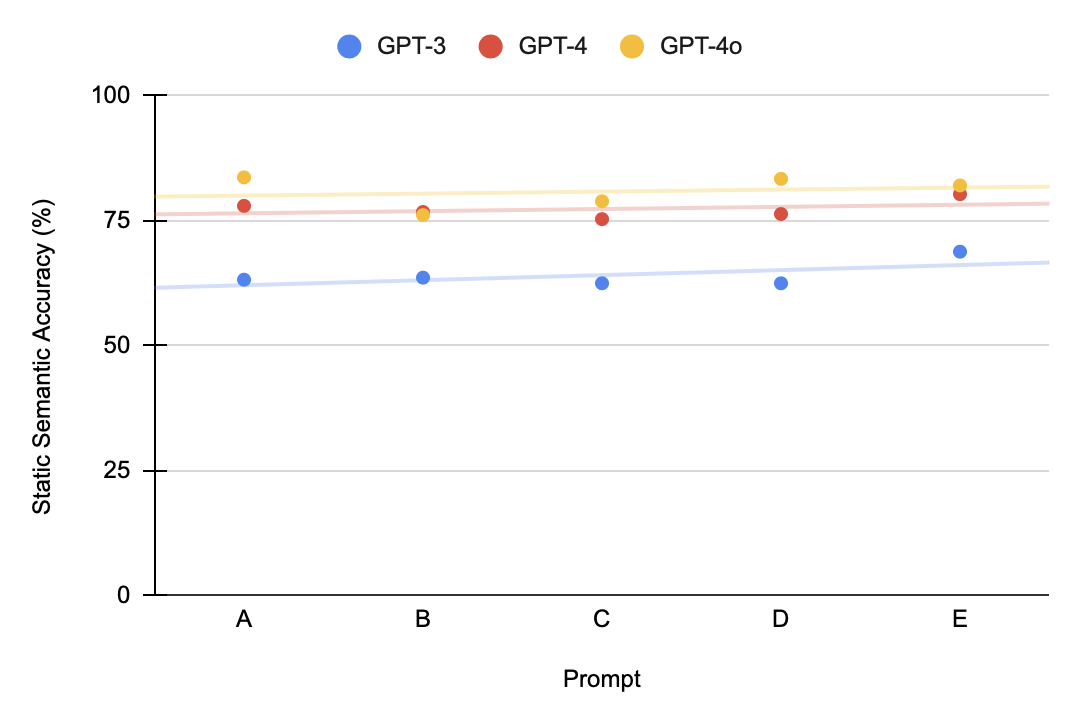}
        \caption{Static Semantic Accuracy.}
        \label{fig:static_semantic_accuracy}
    \end{minipage}

    \vspace{0.3cm}

    \begin{minipage}[b]{0.49\textwidth}
        \centering
        \includegraphics[width=\textwidth]{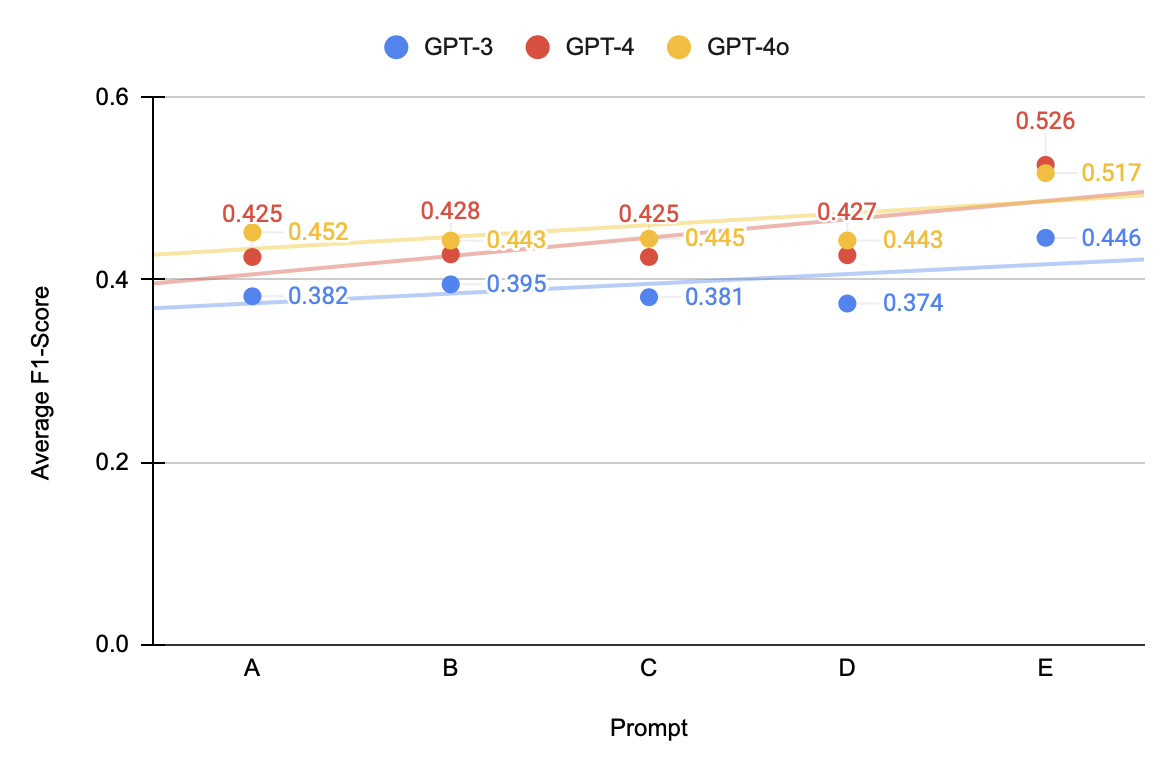}
        \caption{Average ROUGE-L Score.}
        \label{fig:average_rougeL_score}
    \end{minipage}
     \hfill
    \begin{minipage}[b]{0.49\textwidth}
        \centering
        \includegraphics[width=\textwidth]{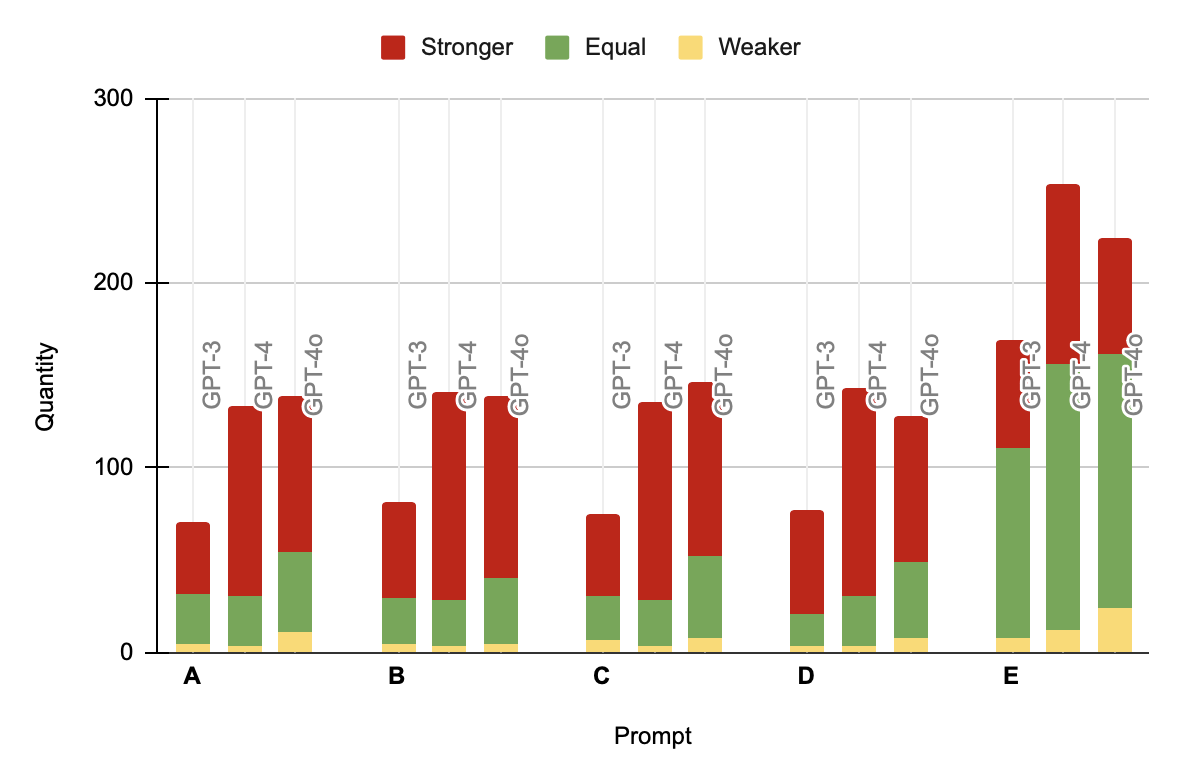}
        \caption{Assertions Equality.}
        \label{fig:weaker_equal_stronger}
    \end{minipage}
    
\end{figure*}

\begin{figure*}[h] 
    \centering
    \begin{minipage}[b]{0.49\textwidth}
        \centering
        \includegraphics[width=\textwidth]{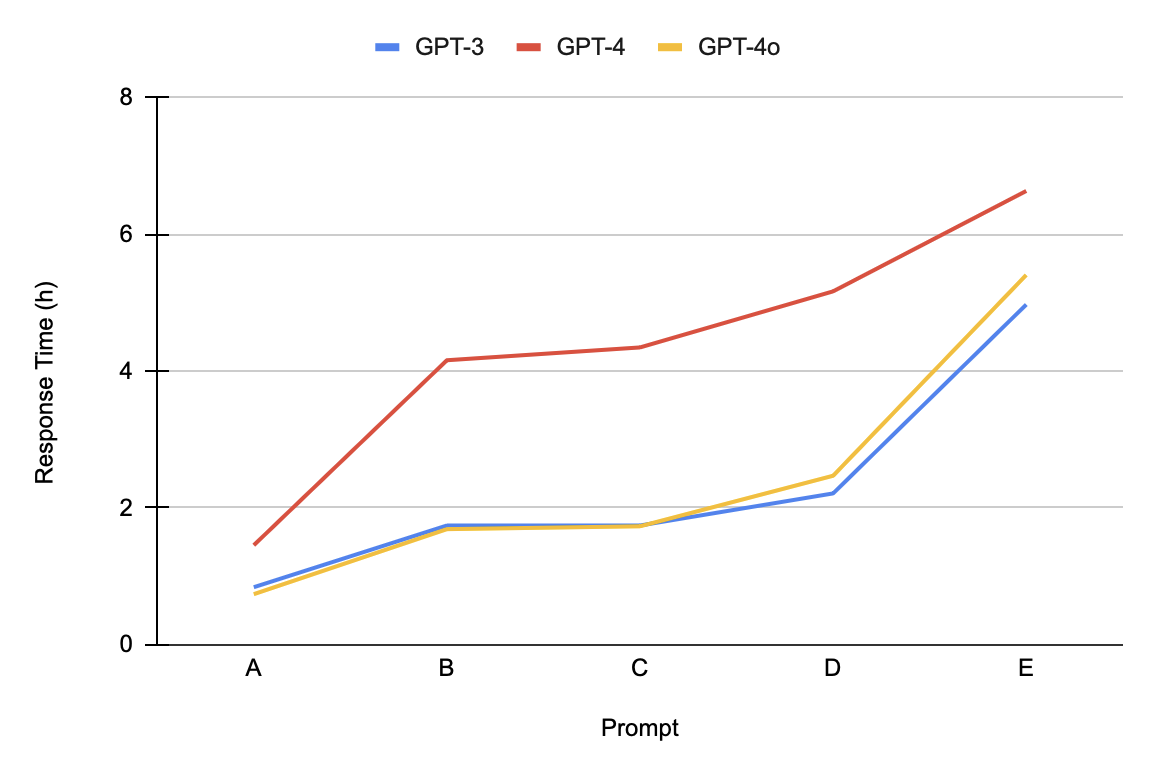}
        \caption{Response Time.}
        \label{fig:response_time}
    \end{minipage}
    \hfill
    \begin{minipage}[b]{0.49\textwidth}
        \centering
        \includegraphics[width=\textwidth]{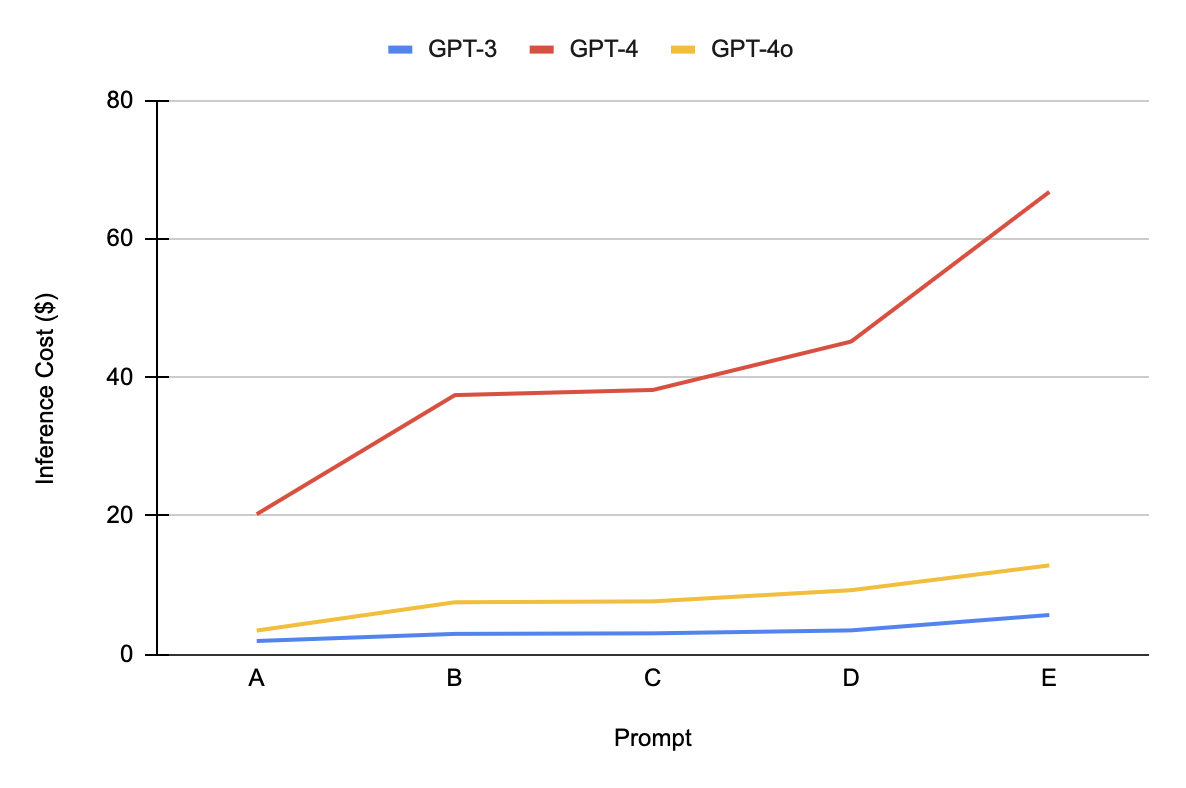}
        \caption{Inference Cost.}
        \label{fig:inference_cost}
    \end{minipage}
\end{figure*}

This section presents the results of our experiments with five distinct prompts conducted using the GPT-3, GPT-4, and GPT-4o models. We selected these models for their diversity in inference time, cost, and, most importantly, accuracy, allowing developers and researchers to choose the model that best suits their specific needs.

Overall, the experiments demonstrate the high accuracy of \tool across all prompts and models, achieving the best when using the GPT-4o model. It also shows the importance of few-shot learning as a key factor in increasing the structural similarity to the original ones. Figures \ref{fig:sne_rate} and \ref{fig:syntactic_accuracy} support this claim, showing that the syntactic accuracy of the generated assertions ranges from 87.5\% to 97.4\%, depending on the prompt chosen by \tool. Regarding static semantic accuracy, results vary between 62.3\% and 83.5\%. Additionally, the experiments reveal a structural similarity of 52.6\%, as measured by the ROUGE-L $F_{1}$-Score for the latest prompt, \textit{E}. We also evaluate \tool's performance in terms of response time and inference cost, as illustrated in Figures \ref{fig:response_time} and \ref{fig:inference_cost}. 

\begin{lstlisting}[style=JavaStyle,showstringspaces=false,tabsize=2,label=lst:parsing_error_sample,caption= Example of Parsing Error (Prompt A / GPT-4).,captionpos=b, float=h, basicstyle=\scriptsize]
@eassert name != null;e@
1. @kpublick@ IRubyObject fetchClassVariable(String name) {
2.     @kreturnk@ getClassVariablesForRead().get(name);
3. }
\end{lstlisting}
    
\begin{enumerate}[wide=0pt]
    \item \textbf{RQ1}: To address this RQ, we refer to Figure \ref{fig:syntactic_accuracy} to evaluate how well \tool generates syntactically correct assertions that can be inserted into a given method without breaking parsability. As shown in the figure, the overall syntactic accuracy ranges from 96.0\% to 97.4\% when using the GPT-4o model, 91.8\% to 96.6\% for GPT-4, and 88.1\% to 90.8\% for GPT-3. The tool performs best with the final prompt, \textit{E}, when using the GPT-4o model. The mean success rate of all experiments across all models for syntactical accuracy according to Figure \ref{fig:syntactic_accuracy} is 93.18\%.

    We next aimed to identify which prompt engineering technique contributes the most to syntactic accuracy by incrementally removing elements from the best-performing prompt, \textit{E}, in an ablation study. This analysis examines the impact of each component on the tool’s syntactic accuracy. Figure \ref{fig:sne_rate} shows that transitioning from prompt \textit{E} to \textit{D} yields the highest increase in SNE rate, with a 3.25\% decline in accuracy, while other prompt modifications in the ablation study had less impact on the SNE rate. Our findings further indicate that including the method's input/output (\textit{C}) consistently increases \tool's SNE rate across models, whereas adding a summary of the method’s external dependencies reduces SNE errors for the GPT-4o model.

    We also analyzed the causes of syntactic errors that prevented successful parsing after assertion insertion. Our analysis revealed that the primary cause of these errors is incorrect line numbers, which render the resulting code unparsable. Listing \ref{lst:parsing_error_sample} illustrates an example where \tool generated line number 1 for inserting an assertion, whereas line 2 would have been correct to assess the nullability of the variable \texttt{name}. Another reason is the LLM's violation of the expected output template, resulting in malformed extracted assertions.\\

    \begin{table}[h]
    \label{box:rq1_summary}
    \begin{tabularx}{0.97\textwidth}{|>{\RaggedRight\arraybackslash}X|}
      \hline
        \tool demonstrates consistent efficiency in generating syntactically correct assertions. The syntactic accuracy ranges from 96.0\% to 97.4\%, performing best with few-shot learning using the GPT-4o model.\\
      \hline
    \end{tabularx}
    \end{table}

    \item \textbf{RQ2}: To address this RQ, we refer to Figure \ref{fig:static_semantic_accuracy} which shows the static semantic accuracy of \tool using different prompt. Overall, \tool performs best when using the GPT-4o model, achieving a maximum accuracy of 83.5\%. The mean success rate of all experiments across all models for static semantic accuracy according to Figure \ref{fig:static_semantic_accuracy} is 73.95\%, while the accuracy ranges from 62.3\% to 83.5\%. Our experiments show that prompt engineering does not directly contribute to a consistent decrease in the SME rate, while the linear trend line increases. Comparing the most naive prompt (\textit{A}) with the most complex prompt (\textit{E}), we conclude a positive impact of prompt engineering when adopting the GPT-3 and GPT-4 models.
  
    Our analysis identified three main causes for generating semantically incorrect assertions:

    \begin{enumerate}
         \item \textbf{Undefined symbol}: This error occurs when \tool generates assertions containing an undefined symbol (variable, method, or class). 
         In these cases, \tool incorrectly predicts a location for an assertion using one or more symbols before those symbols are declared. 
        \item \textbf{Unreachable statement}: This error occurs when \tool generates an assertion to be inserted after a \texttt{return} statement, preventing successful compilation. 
        Similar to the previous error, incorrect line numbers interpreted by \tool lead to generating assertions at locations that will never be executed.
        \item \textbf{Bad operand types}: This error occurs due to a mismatch between operators and operands in a generated assertion.
        Listing \ref{lst:mot_naive_code_b} illustrates such errors, where the integer variables \texttt{AA} and \texttt{BBBBBBBB} are incorrectly compared with null, though this is illegal within the integer domain. 
    \end{enumerate}
    \hfill \break
    \vspace{-0.5cm}
    \begin{table}[h]
    \label{box:rq2_summary}
    \begin{tabularx}{0.97\textwidth}{|>{\RaggedRight\arraybackslash}X|}
      \hline
        \tool is capable of generating assertions with a high static semantic accuracy ranging from 62.3\% to 83.5\%, showing an increasing trend with prompt engineering while performing best with the GPT-4o model.\\
      \hline
    \end{tabularx}
    \end{table}
    
    \item \textbf{RQ3}: To answer this RQ, we refer to Figure \ref{fig:average_rougeL_score} which shows the average ROUGE-L score for each prompt to measure structural similarity. In Figure \ref{fig:average_rougeL_score}, the average ROUGE-L score for each prompt along with its trend line is shown, ranging from 0.374 for GPT-3 up to 0.526 for GPT-4. The figure shows that \tool performs best (overall) when using the GPT-4o model, generating the most similar assertions. The trend line for all three models shows a general increase, meaning that the model overall tends to generate assertions closer to the originals as it receives more information about the model. Few-shot learning shows a positive effect, increasing the ROUGE-L score by 0.09 with the GPT-4 model, and 0.07 with both GPT-3 and GPT-4o models. Thus, while exact matches are not guaranteed, \tool achieves the best ROUGE-L similarity of 0.52 and 0.51 with the GPT-4 and GPT-4o models, respectively.

    Regarding the comparative strength of generated assertions, Figure \ref{fig:weaker_equal_stronger} demonstrates that among the instances of original or generated assertions where one set is a subset of the other, the generated assertions are either equal to or stronger than the originals. In other words, ground truth assertions are often fully represented by \tool. Our experiments also indicate that few-shot learning enhances \tool's ability to generate equal sets of assertions compared to adding other prompt elements. Another observation is that the GPT-4 model generates stronger assertions more frequently than the other two models. 
    
    \begin{table}[h]
    \label{box:rq3_summary}
    \begin{tabularx}{0.97\textwidth}{|>{\RaggedRight\arraybackslash}X|}
      \hline
        \tool can generate structurally similar assertions compared to the original assertions, with the average ROUGE-L scores ranging from 0.374 to 0.526, showing an upward trend with prompt engineering and performing best with few-shot learning.\\
      \hline
    \end{tabularx}
    \end{table}
\end{enumerate}

In terms of inference time and response time for the given dataset during conducting experiments, Figures \ref{fig:response_time} and \ref{fig:inference_cost} indicate that enriching the prompt with additional context generally leads to increases in cost and time. These rises are attributed to the complexity of prompt construction, which is particularly evident when \tool uses the GPT-4 model. When considered alongside the previous figures, this information provides developers with greater insight into the affordability of each configuration for generating production assertions.
\section{Threats To Validity}
\label{sec:threatstovalidity}

Our research represents a preliminary effort to harness the capabilities of LLMs in generating production assertions. However, certain factors may jeopardize the validity of our findings. First, since the \tool leverages models' APIs, the quality of responses—such as malformed outputs or hallucinations—is directly tied to the performance of these models. We tried to mitigate such issues through the \textit{postprocessor} component to filter malformed outputs. Additionally, as \tool is designed to work with other LLMs, any internal problems or instabilities in those models, such as overfitting, underfitting, or biased results, could threaten the validity of our findings.

Another potential threat arises from the quality of the input datasets, which may affect not only the candidate methods for which \tool generates assertions but also the few-shot learning dataset and, consequently, prompt engineering. 
We tried to mitigate this by selecting mature and popular repositories with some filters, as discussed in Section \ref{sec:experiment}.

\section{Related Work}
\label{sec:relatedwork}

We categorize related research works into four main areas. 
These works propose different assertion generation approaches that mainly target unit tests, while ours focuses on production assertions mainly used for comprehension and debugging.

\subsection{Dynamic approaches}
Among the most notable contributions in this area is Daikon\cite{Ernst2000_PhD}, recognized as a leading tool in dynamic invariant detection. Daikon observes program executions over a test set and monitors variables to infer consistent relationships through statistical analysis and pattern recognition, resulting in identifying likely invariants. It also extends invariant detection to pointer-based data structure and linearizes implicit collections into arrays, making it a powerful research work compared to its counterparts. Another research work in this area is DIG \cite{DIG}, a Dynamic Invariant Generator, analyzing C/C++ program executions to infer polynomial and array invariants using abstract interpretation and constraint solving. While both works have often focused on program execution, our research does not solely rely on execution except for running unit tests. Compared to DIG, \tool offers a broader range of assertions beyond polynomials and array invariants. Moreover, switching between various LLMs makes our approach more adaptable and effective.

\subsection{Static approaches}
These approaches examine source code over the possible runtime states and generate unit tests accordingly. A common approach within this category is generating the unit tests randomly. Randoop\cite{10.1145/1297846.1297902} generates random sequences of Java method calls on objects and checks for any errors occurring. If any error happens, a unit test is created by running these invocation sequences. Another technique within this area is search-based testing, which uses efficient meta-heuristic search algorithms for test generation. For instance,  Evosuite\cite{10.1145/2025113.2025179} uses this technique and relies on an evolutionary approach based on a genetic algorithm to generate unit tests, targeting code coverage. A significant weakness and criticism of the static approaches are the low code quality and low understandability of the generated test cases\cite{10.1145/1297846.1297902}. Moreover, they entirely create unit tests while the \tool generates assertions for production code.

\subsection{Learning-based approaches}
Neural networks and LLMs have been used to generate testing oracles. Yu and Lou \cite{yu2019automated} propose an Information Retrieval-based \cite{IR} approach that integrates Word2Vec\cite{casellesdupré2018word2vec} and a bidirectional RNN to enhance the effectiveness of ATLAS\cite{ATLAS}, a deep learning based approach to generate assertions based on provided unit tests and refine assertion adaptation. It consists of two parts: IR-based assertion retrieval extracts the most similar focal tests using query-based search and adopts their corresponding assertions as output. Retrieved-assertion adaptation evaluates and identifies discrepancies between the tokens, focusing on adapting to align with the test context. However, compared to \tool, their approach doesn't check assertions' compilability and only concentrates on adapting existing ones. Wang et al. \cite{10.1145/3650105.3652293} fine-tuned the GraphCodeBERT model \cite{guo2021graphcodebertpretrainingcoderepresentations} using high-dimensional vectors that represent encoded source code information, along with sequences of testing assertions. Their approach demonstrated promising results in terms of the number and types of correct testing assertions generated, as well as the ability to replicate the exact assert statements written by developers. Schafer and Nadi present an approach called TESTPILOT \cite{schäfer2023empirical} that employs the Codex model on Mocha\cite{mocha} framework, having a substantial statement coverage directly through prompts, complemented by an adaptive mechanism to refine failed tests iteratively. Overall, \tool differs by handling overloaded functions using metadata, and working with Java as a statically typed language with adherence to compilation, while TESTPILOT focuses on JavaScript as a dynamically typed language.

\subsection{Code-generation approaches}
This category focuses on leveraging LLMs for code generation and summarizing, making them less adhered to generating assertions.
Liventsev and Grishina present \textit{Synthesize, Execute, Inspect, Debug, Replace (SEIDR)} approach \cite{fullyautonomousprogrammingwithllm2023} designed for programming tasks defined by textual descriptions and I/O examples. This research enhances program synthesis by integrating debugging for refining generated programs that fail tests, having a synergy between LLMs and an iterative debug cycle to optimize code generation and repair. Regarding code summary generation, MacNeil and Tran explore integrating code explanations generated by LLMs into an interactive e-book for web software development education \cite{experience_from_using_code_explanation}. This study investigates the engagement and perceived utility of LLMs among students, showcasing the capability of LLMs in generating code summaries. Bhattacharya and Chakraborty evaluate the effectiveness of various LLMs in generating natural-language summaries for code snippets \cite{exploring_llm_for_code_explanation}. the study demonstrates that code-specific LLMs outperform generic models using token-based (BLEU scores) and semantics-based metrics (CodeBERT embeddings).

\section{Conclusion}
\label{sec:conclusion}

This research presents a novel approach, called \tool, using prompt-engineered Large Language Models to automate the generation of production assertions, distinct from past research focused on test assertions. It evaluates the generated assertions' syntactic and static semantic accuracy and compares their structural similarity to assertions written by developers. Our experiments on large, mature Java repositories show \tool's effectiveness when using the GPT-4o model. They also show the effectiveness of few-shot learning in increasing the accuracy of the tool when using the GPT-4 and GPT-3 models. In the future, we plan to explore open-source LLMs and extend \tool to support additional build tools and programming languages. We also plan to conduct a user study to assess the quality of the generated assertions from the perspective of code maintainers.
\section{Data Availability}
\label{sec:data_availability}

All artifacts including the source codes and experiments' results are available online at \url{https://anonymous.4open.science/r/AssertionGenerationArxiv/README.md}.

\bibliographystyle{acm_reference_format}
\bibliography{ref}


\begin{thebibliography}{42}


\ifx \showCODEN    \undefined \def \showCODEN     #1{\unskip}     \fi
\ifx \showDOI      \undefined \def \showDOI       #1{#1}\fi
\ifx \showISBNx    \undefined \def \showISBNx     #1{\unskip}     \fi
\ifx \showISBNxiii \undefined \def \showISBNxiii  #1{\unskip}     \fi
\ifx \showISSN     \undefined \def \showISSN      #1{\unskip}     \fi
\ifx \showLCCN     \undefined \def \showLCCN      #1{\unskip}     \fi
\ifx \shownote     \undefined \def \shownote      #1{#1}          \fi
\ifx \showarticletitle \undefined \def \showarticletitle #1{#1}   \fi
\ifx \showURL      \undefined \def \showURL       {\relax}        \fi
\providecommand\bibfield[2]{#2}
\providecommand\bibinfo[2]{#2}
\providecommand\natexlab[1]{#1}
\providecommand\showeprint[2][]{arXiv:#2}

\bibitem[Bhattacharya et~al\mbox{.}(2023)]%
        {exploring_llm_for_code_explanation}
\bibfield{author}{\bibinfo{person}{Paheli Bhattacharya}, \bibinfo{person}{Manojit Chakraborty}, \bibinfo{person}{Kartheek N S~N Palepu}, \bibinfo{person}{Vikas Pandey}, \bibinfo{person}{Ishan Dindorkar}, \bibinfo{person}{Rakesh Rajpurohit}, {and} \bibinfo{person}{Rishabh Gupta}.} \bibinfo{year}{2023}\natexlab{}.
\newblock \bibinfo{title}{Exploring Large Language Models for Code Explanation}.
\newblock
\newblock
\showeprint[arxiv]{2310.16673}~[cs.SE]


\bibitem[Cambaz and Zhang(2024)]%
        {use_of_ai_driven_code_generation_models}
\bibfield{author}{\bibinfo{person}{Doga Cambaz} {and} \bibinfo{person}{Xiaoling Zhang}.} \bibinfo{year}{2024}\natexlab{}.
\newblock \showarticletitle{Use of AI-driven Code Generation Models in Teaching and Learning Programming: a Systematic Literature Review}. In \bibinfo{booktitle}{\emph{Proceedings of the 55th ACM Technical Symposium on Computer Science Education V. 1}} (<conf-loc>, <city>Portland</city>, <state>OR</state>, <country>USA</country>, </conf-loc>) \emph{(\bibinfo{series}{SIGCSE 2024})}. \bibinfo{publisher}{Association for Computing Machinery}, \bibinfo{address}{New York, NY, USA}, \bibinfo{pages}{172–178}.
\newblock
\showISBNx{9798400704239}
\urldef\tempurl%
\url{https://doi.org/10.1145/3626252.3630958}
\showDOI{\tempurl}


\bibitem[Caselles-Dupré et~al\mbox{.}(2018)]%
        {casellesdupré2018word2vec}
\bibfield{author}{\bibinfo{person}{Hugo Caselles-Dupré}, \bibinfo{person}{Florian Lesaint}, {and} \bibinfo{person}{Jimena Royo-Letelier}.} \bibinfo{year}{2018}\natexlab{}.
\newblock \bibinfo{title}{Word2Vec applied to Recommendation: Hyperparameters Matter}.
\newblock
\newblock
\showeprint[arxiv]{1804.04212}~[cs.IR]


\bibitem[Chang et~al\mbox{.}(2024)]%
        {10.1145/3641289}
\bibfield{author}{\bibinfo{person}{Yupeng Chang}, \bibinfo{person}{Xu Wang}, \bibinfo{person}{Jindong Wang}, \bibinfo{person}{Yuan Wu}, \bibinfo{person}{Linyi Yang}, \bibinfo{person}{Kaijie Zhu}, \bibinfo{person}{Hao Chen}, \bibinfo{person}{Xiaoyuan Yi}, \bibinfo{person}{Cunxiang Wang}, \bibinfo{person}{Yidong Wang}, \bibinfo{person}{Wei Ye}, \bibinfo{person}{Yue Zhang}, \bibinfo{person}{Yi Chang}, \bibinfo{person}{Philip~S. Yu}, \bibinfo{person}{Qiang Yang}, {and} \bibinfo{person}{Xing Xie}.} \bibinfo{year}{2024}\natexlab{}.
\newblock \showarticletitle{A Survey on Evaluation of Large Language Models}.
\newblock \bibinfo{journal}{\emph{ACM Trans. Intell. Syst. Technol.}} \bibinfo{volume}{15}, \bibinfo{number}{3}, Article \bibinfo{articleno}{39} (\bibinfo{date}{mar} \bibinfo{year}{2024}), \bibinfo{numpages}{45}~pages.
\newblock
\showISSN{2157-6904}
\urldef\tempurl%
\url{https://doi.org/10.1145/3641289}
\showDOI{\tempurl}


\bibitem[Chen et~al\mbox{.}(2023)]%
        {chen2024unleashingpotentialpromptengineering}
\bibfield{author}{\bibinfo{person}{Banghao Chen}, \bibinfo{person}{Zhaofeng Zhang}, \bibinfo{person}{Nicolas Langrené}, {and} \bibinfo{person}{Shengxin Zhu}.} \bibinfo{year}{2023}\natexlab{}.
\newblock \bibinfo{title}{Unleashing the potential of prompt engineering in Large Language Models: a comprehensive review}.
\newblock
\newblock


\bibitem[Company(2013)]%
        {sourceGraph}
\bibfield{author}{\bibinfo{person}{Sourcegraph Company}.} \bibinfo{year}{2013}\natexlab{}.
\newblock \bibinfo{booktitle}{\emph{Repositories dataset extraction source}}.
\newblock
\urldef\tempurl%
\url{https://sourcegraph.com/}
\showURL{%
Retrieved Aug 20, 2023 from \tempurl}


\bibitem[Dinella et~al\mbox{.}(2022a)]%
        {TOGA_2022}
\bibfield{author}{\bibinfo{person}{Elizabeth Dinella}, \bibinfo{person}{Gabriel Ryan}, \bibinfo{person}{Todd Mytkowicz}, {and} \bibinfo{person}{Shuvendu~K. Lahiri}.} \bibinfo{year}{2022}\natexlab{a}.
\newblock \showarticletitle{TOGA: a neural method for test oracle generation}. In \bibinfo{booktitle}{\emph{Proceedings of the 44th International Conference on Software Engineering}} \emph{(\bibinfo{series}{ICSE ’22})}. \bibinfo{publisher}{ACM}.
\newblock
\urldef\tempurl%
\url{https://doi.org/10.1145/3510003.3510141}
\showDOI{\tempurl}


\bibitem[Dinella et~al\mbox{.}(2022b)]%
        {dinella}
\bibfield{author}{\bibinfo{person}{Elizabeth Dinella}, \bibinfo{person}{Gabriel Ryan}, \bibinfo{person}{Todd Mytkowicz}, {and} \bibinfo{person}{Shuvendu~K. Lahiri}.} \bibinfo{year}{2022}\natexlab{b}.
\newblock \showarticletitle{TOGA: a neural method for test oracle generation}. In \bibinfo{booktitle}{\emph{Proceedings of the 44th International Conference on Software Engineering}} (Pittsburgh, Pennsylvania) \emph{(\bibinfo{series}{ICSE '22})}. \bibinfo{publisher}{Association for Computing Machinery}, \bibinfo{address}{New York, NY, USA}, \bibinfo{pages}{2130–2141}.
\newblock
\showISBNx{9781450392211}
\urldef\tempurl%
\url{https://doi.org/10.1145/3510003.3510141}
\showDOI{\tempurl}


\bibitem[Dinh et~al\mbox{.}(2011)]%
        {5948597}
\bibfield{author}{\bibinfo{person}{Minh~Ngoc Dinh}, \bibinfo{person}{David Abramson}, \bibinfo{person}{Donny Kurniawan}, \bibinfo{person}{Chao Jin}, \bibinfo{person}{Bob Moench}, {and} \bibinfo{person}{Luiz DeRose}.} \bibinfo{year}{2011}\natexlab{}.
\newblock \showarticletitle{Assertion Based Parallel Debugging}. In \bibinfo{booktitle}{\emph{2011 11th IEEE/ACM International Symposium on Cluster, Cloud and Grid Computing}}. \bibinfo{pages}{63--72}.
\newblock
\urldef\tempurl%
\url{https://doi.org/10.1109/CCGrid.2011.44}
\showDOI{\tempurl}


\bibitem[Elkarablieh et~al\mbox{.}(2007)]%
        {10.1145/1321631.1321643}
\bibfield{author}{\bibinfo{person}{Bassem Elkarablieh}, \bibinfo{person}{Ivan Garcia}, \bibinfo{person}{Yuk~Lai Suen}, {and} \bibinfo{person}{Sarfraz Khurshid}.} \bibinfo{year}{2007}\natexlab{}.
\newblock \showarticletitle{Assertion-based repair of complex data structures}. \bibinfo{publisher}{Association for Computing Machinery}, \bibinfo{address}{New York, NY, USA}.
\newblock
\showISBNx{9781595938824}
\urldef\tempurl%
\url{https://doi.org/10.1145/1321631.1321643}
\showDOI{\tempurl}


\bibitem[Ernst(2000)]%
        {Ernst2000_PhD}
\bibfield{author}{\bibinfo{person}{Michael~D. Ernst}.} \bibinfo{year}{2000}\natexlab{}.
\newblock \emph{\bibinfo{title}{Dynamically Discovering Likely Program Invariants}}.
\newblock {Ph.D.} \bibinfo{school}{University of Washington Department of Computer Science and Engineering}, \bibinfo{address}{Seattle, Washington}.
\newblock


\bibitem[Fraser and Arcuri(2011)]%
        {10.1145/2025113.2025179}
\bibfield{author}{\bibinfo{person}{Gordon Fraser} {and} \bibinfo{person}{Andrea Arcuri}.} \bibinfo{year}{2011}\natexlab{}.
\newblock \showarticletitle{EvoSuite: automatic test suite generation for object-oriented software}. In \bibinfo{booktitle}{\emph{Proceedings of the 19th ACM SIGSOFT Symposium and the 13th European Conference on Foundations of Software Engineering}} (Szeged, Hungary) \emph{(\bibinfo{series}{ESEC/FSE '11})}. \bibinfo{publisher}{Association for Computing Machinery}, \bibinfo{address}{New York, NY, USA}, \bibinfo{pages}{416–419}.
\newblock
\showISBNx{9781450304436}
\urldef\tempurl%
\url{https://doi.org/10.1145/2025113.2025179}
\showDOI{\tempurl}


\bibitem[Guo et~al\mbox{.}(2020)]%
        {guo2021graphcodebertpretrainingcoderepresentations}
\bibfield{author}{\bibinfo{person}{Daya Guo}, \bibinfo{person}{Shuo Ren}, \bibinfo{person}{Shuai Lu}, \bibinfo{person}{Zhangyin Feng}, \bibinfo{person}{Duyu Tang}, \bibinfo{person}{Shujie Liu}, \bibinfo{person}{Long Zhou}, \bibinfo{person}{Nan Duan}, \bibinfo{person}{Jian Yin}, \bibinfo{person}{Daxin Jiang}, {and} \bibinfo{person}{Ming Zhou}.} \bibinfo{year}{2020}\natexlab{}.
\newblock \bibinfo{title}{GraphCodeBERT: Pre-training Code Representations with Data Flow}.
\newblock
\newblock
\urldef\tempurl%
\url{https://doi.org/10.48550/arXiv.2009.08366}
\showDOI{\tempurl}


\bibitem[Hambarde and Proença(2023)]%
        {IR}
\bibfield{author}{\bibinfo{person}{Kailash~A. Hambarde} {and} \bibinfo{person}{Hugo Proença}.} \bibinfo{year}{2023}\natexlab{}.
\newblock \showarticletitle{Information Retrieval: Recent Advances and Beyond}.
\newblock \bibinfo{journal}{\emph{IEEE Access}}  \bibinfo{volume}{11} (\bibinfo{year}{2023}), \bibinfo{pages}{76581–76604}.
\newblock
\showISSN{2169-3536}
\urldef\tempurl%
\url{https://doi.org/10.1109/access.2023.3295776}
\showDOI{\tempurl}


\bibitem[Kang et~al\mbox{.}(2023)]%
        {kang2023largelanguagemodelsfewshot}
\bibfield{author}{\bibinfo{person}{Sungmin Kang}, \bibinfo{person}{Juyeon Yoon}, {and} \bibinfo{person}{Shin Yoo}.} \bibinfo{year}{2023}\natexlab{}.
\newblock \showarticletitle{Large Language Models are Few-Shot Testers: Exploring LLM-Based General Bug Reproduction}. In \bibinfo{booktitle}{\emph{Proceedings of the 45th International Conference on Software Engineering}} (Melbourne, Victoria, Australia) \emph{(\bibinfo{series}{ICSE '23})}. \bibinfo{publisher}{IEEE Press}, \bibinfo{pages}{2312–2323}.
\newblock
\showISBNx{9781665457019}
\urldef\tempurl%
\url{https://doi.org/10.1109/ICSE48619.2023.00194}
\showDOI{\tempurl}


\bibitem[Korel and Al-Yami(1996)]%
        {assertion_oriented_reference}
\bibfield{author}{\bibinfo{person}{Bogdan Korel} {and} \bibinfo{person}{Ali~M. Al-Yami}.} \bibinfo{year}{1996}\natexlab{}.
\newblock \showarticletitle{Assertion-oriented automated test data generation}. In \bibinfo{booktitle}{\emph{Proceedings of the 18th International Conference on Software Engineering}} (Berlin, Germany) \emph{(\bibinfo{series}{ICSE '96})}. \bibinfo{publisher}{IEEE Computer Society}, \bibinfo{address}{USA}, \bibinfo{pages}{71–80}.
\newblock
\showISBNx{0818672463}


\bibitem[Kudrjavets et~al\mbox{.}(2006)]%
        {Kudrjavets_paper}
\bibfield{author}{\bibinfo{person}{Gunnar Kudrjavets}, \bibinfo{person}{Nachiappan Nagappan}, {and} \bibinfo{person}{Thomas Ball}.} \bibinfo{year}{2006}\natexlab{}.
\newblock \showarticletitle{Assessing the Relationship between Software Assertions and Faults: An Empirical Investigation}.
\newblock \bibinfo{journal}{\emph{Proceedings - International Symposium on Software Reliability Engineering, ISSRE}}, \bibinfo{pages}{204--212}.
\newblock
\urldef\tempurl%
\url{https://doi.org/10.1109/ISSRE.2006.14}
\showDOI{\tempurl}


\bibitem[Leinonen et~al\mbox{.}(2023)]%
        {comparing_code_explanations}
\bibfield{author}{\bibinfo{person}{Juho Leinonen}, \bibinfo{person}{Paul Denny}, \bibinfo{person}{Stephen MacNeil}, \bibinfo{person}{Sami Sarsa}, \bibinfo{person}{Seth Bernstein}, \bibinfo{person}{Joanne Kim}, \bibinfo{person}{Andrew Tran}, {and} \bibinfo{person}{Arto Hellas}.} \bibinfo{year}{2023}\natexlab{}.
\newblock \bibinfo{title}{Comparing Code Explanations Created by Students and Large Language Models}.
\newblock
\newblock
\showeprint[arxiv]{2304.03938}~[cs.CY]


\bibitem[Lin(2004)]%
        {lin2004rouge}
\bibfield{author}{\bibinfo{person}{Chin-Yew Lin}.} \bibinfo{year}{2004}\natexlab{}.
\newblock \showarticletitle{Rouge: A package for automatic evaluation of summaries}. In \bibinfo{booktitle}{\emph{Text summarization branches out}}. \bibinfo{pages}{74--81}.
\newblock


\bibitem[Liventsev et~al\mbox{.}(2023)]%
        {fullyautonomousprogrammingwithllm2023}
\bibfield{author}{\bibinfo{person}{Vadim Liventsev}, \bibinfo{person}{Anastasiia Grishina}, \bibinfo{person}{Aki Härmä}, {and} \bibinfo{person}{Leon Moonen}.} \bibinfo{year}{2023}\natexlab{}.
\newblock \showarticletitle{Fully Autonomous Programming with Large Language Models}. In \bibinfo{booktitle}{\emph{Proceedings of the Genetic and Evolutionary Computation Conference}}. \bibinfo{publisher}{{ACM}}.
\newblock
\urldef\tempurl%
\url{https://doi.org/10.1145/3583131.3590481}
\showDOI{\tempurl}


\bibitem[MacNeil et~al\mbox{.}(2023)]%
        {experience_from_using_code_explanation}
\bibfield{author}{\bibinfo{person}{Stephen MacNeil}, \bibinfo{person}{Andrew Tran}, \bibinfo{person}{Arto Hellas}, \bibinfo{person}{Joanne Kim}, \bibinfo{person}{Sami Sarsa}, \bibinfo{person}{Paul Denny}, \bibinfo{person}{Seth Bernstein}, {and} \bibinfo{person}{Juho Leinonen}.} \bibinfo{year}{2023}\natexlab{}.
\newblock \showarticletitle{Experiences from Using Code Explanations Generated by Large Language Models in a Web Software Development E-Book}. In \bibinfo{booktitle}{\emph{Proceedings of the 54th ACM Technical Symposium on Computer Science Education V. 1}} (<conf-loc>, <city>Toronto ON</city>, <country>Canada</country>, </conf-loc>) \emph{(\bibinfo{series}{SIGCSE 2023})}. \bibinfo{publisher}{Association for Computing Machinery}, \bibinfo{address}{New York, NY, USA}, \bibinfo{pages}{931–937}.
\newblock
\showISBNx{9781450394314}
\urldef\tempurl%
\url{https://doi.org/10.1145/3545945.3569785}
\showDOI{\tempurl}


\bibitem[MacNeil et~al\mbox{.}(2022)]%
        {generating_diverse_code_explanations}
\bibfield{author}{\bibinfo{person}{Stephen MacNeil}, \bibinfo{person}{Andrew Tran}, \bibinfo{person}{Dan Mogil}, \bibinfo{person}{Seth Bernstein}, \bibinfo{person}{Erin Ross}, {and} \bibinfo{person}{Ziheng Huang}.} \bibinfo{year}{2022}\natexlab{}.
\newblock \showarticletitle{Generating Diverse Code Explanations using the GPT-3 Large Language Model}. In \bibinfo{booktitle}{\emph{Proceedings of the 2022 ACM Conference on International Computing Education Research - Volume 2}} (Lugano and Virtual Event, Switzerland) \emph{(\bibinfo{series}{ICER '22})}. \bibinfo{publisher}{Association for Computing Machinery}, \bibinfo{address}{New York, NY, USA}, \bibinfo{pages}{37–39}.
\newblock
\showISBNx{9781450391955}
\urldef\tempurl%
\url{https://doi.org/10.1145/3501709.3544280}
\showDOI{\tempurl}


\bibitem[{Mocha maintainers}(2023)]%
        {mocha}
\bibfield{author}{\bibinfo{person}{{Mocha maintainers}}.} \bibinfo{year}{2023}\natexlab{}.
\newblock \bibinfo{title}{Mocha - the fun, simple, flexible JavaScript test framework}.
\newblock
\newblock
\urldef\tempurl%
\url{https://mochajs.org/}
\showURL{%
\tempurl}


\bibitem[Nguyen et~al\mbox{.}(2014)]%
        {DIG}
\bibfield{author}{\bibinfo{person}{Thanhvu Nguyen}, \bibinfo{person}{Deepak Kapur}, \bibinfo{person}{Westley Weimer}, {and} \bibinfo{person}{Stephanie Forrest}.} \bibinfo{year}{2014}\natexlab{}.
\newblock \showarticletitle{DIG: A Dynamic Invariant Generator for Polynomial and Array Invariants}.
\newblock \bibinfo{journal}{\emph{ACM Transactions on Software Engineering and Methodology}}  \bibinfo{volume}{23} (\bibinfo{date}{09} \bibinfo{year}{2014}).
\newblock
\urldef\tempurl%
\url{https://doi.org/10.1145/2556782}
\showDOI{\tempurl}


\bibitem[Pacheco and Ernst(2007)]%
        {10.1145/1297846.1297902}
\bibfield{author}{\bibinfo{person}{Carlos Pacheco} {and} \bibinfo{person}{Michael~D. Ernst}.} \bibinfo{year}{2007}\natexlab{}.
\newblock \showarticletitle{Randoop: feedback-directed random testing for Java}. In \bibinfo{booktitle}{\emph{Companion to the 22nd ACM SIGPLAN Conference on Object-Oriented Programming Systems and Applications Companion}} (Montreal, Quebec, Canada) \emph{(\bibinfo{series}{OOPSLA '07})}. \bibinfo{publisher}{Association for Computing Machinery}, \bibinfo{address}{New York, NY, USA}, \bibinfo{pages}{815–816}.
\newblock
\showISBNx{9781595938657}
\urldef\tempurl%
\url{https://doi.org/10.1145/1297846.1297902}
\showDOI{\tempurl}


\bibitem[Parnami and Lee(2022)]%
        {parnami2022learningexamplessummaryapproaches}
\bibfield{author}{\bibinfo{person}{Archit Parnami} {and} \bibinfo{person}{Minwoo Lee}.} \bibinfo{year}{2022}\natexlab{}.
\newblock \bibinfo{title}{Learning from Few Examples: A Summary of Approaches to Few-Shot Learning}.
\newblock
\newblock
\urldef\tempurl%
\url{https://doi.org/10.48550/arXiv.2203.04291}
\showDOI{\tempurl}


\bibitem[Reynolds and McDonell(2021)]%
        {reynolds2021promptprogramminglargelanguage}
\bibfield{author}{\bibinfo{person}{Laria Reynolds} {and} \bibinfo{person}{Kyle McDonell}.} \bibinfo{year}{2021}\natexlab{}.
\newblock \showarticletitle{Prompt Programming for Large Language Models: Beyond the Few-Shot Paradigm}. In \bibinfo{booktitle}{\emph{Extended Abstracts of the 2021 CHI Conference on Human Factors in Computing Systems}} (Yokohama, Japan) \emph{(\bibinfo{series}{CHI EA '21})}. \bibinfo{publisher}{Association for Computing Machinery}, \bibinfo{address}{New York, NY, USA}, Article \bibinfo{articleno}{314}, \bibinfo{numpages}{7}~pages.
\newblock
\showISBNx{9781450380959}
\urldef\tempurl%
\url{https://doi.org/10.1145/3411763.3451760}
\showDOI{\tempurl}


\bibitem[Rosenblum(1995)]%
        {341844}
\bibfield{author}{\bibinfo{person}{D.S. Rosenblum}.} \bibinfo{year}{1995}\natexlab{}.
\newblock \showarticletitle{A practical approach to programming with assertions}.
\newblock \bibinfo{journal}{\emph{IEEE Transactions on Software Engineering}} \bibinfo{volume}{21}, \bibinfo{number}{1} (\bibinfo{year}{1995}), \bibinfo{pages}{19--31}.
\newblock
\urldef\tempurl%
\url{https://doi.org/10.1109/32.341844}
\showDOI{\tempurl}


\bibitem[Rosenblum(1992)]%
        {10.1145/143062.143098}
\bibfield{author}{\bibinfo{person}{David~S. Rosenblum}.} \bibinfo{year}{1992}\natexlab{}.
\newblock \showarticletitle{Towards a method of programming with assertions}. In \bibinfo{booktitle}{\emph{Proceedings of the 14th International Conference on Software Engineering}} (Melbourne, Australia) \emph{(\bibinfo{series}{ICSE '92})}. \bibinfo{publisher}{Association for Computing Machinery}, \bibinfo{address}{New York, NY, USA}, \bibinfo{pages}{92–104}.
\newblock
\showISBNx{0897915046}
\urldef\tempurl%
\url{https://doi.org/10.1145/143062.143098}
\showDOI{\tempurl}


\bibitem[Sarsa et~al\mbox{.}(2022)]%
        {automatic_generation_of_programming}
\bibfield{author}{\bibinfo{person}{Sami Sarsa}, \bibinfo{person}{Paul Denny}, \bibinfo{person}{Arto Hellas}, {and} \bibinfo{person}{Juho Leinonen}.} \bibinfo{year}{2022}\natexlab{}.
\newblock \showarticletitle{Automatic Generation of Programming Exercises and Code Explanations Using Large Language Models}. In \bibinfo{booktitle}{\emph{Proceedings of the 2022 ACM Conference on International Computing Education Research - Volume 1}} \emph{(\bibinfo{series}{ICER 2022})}. \bibinfo{publisher}{ACM}.
\newblock
\urldef\tempurl%
\url{https://doi.org/10.1145/3501385.3543957}
\showDOI{\tempurl}


\bibitem[Schäfer et~al\mbox{.}(2023)]%
        {schäfer2023empirical}
\bibfield{author}{\bibinfo{person}{Max Schäfer}, \bibinfo{person}{Sarah Nadi}, \bibinfo{person}{Aryaz Eghbali}, {and} \bibinfo{person}{Frank Tip}.} \bibinfo{year}{2023}\natexlab{}.
\newblock \bibinfo{title}{An Empirical Evaluation of Using Large Language Models for Automated Unit Test Generation}.
\newblock
\newblock
\showeprint[arxiv]{2302.06527}~[cs.SE]


\bibitem[Terragni et~al\mbox{.}(2020)]%
        {10.1145/3368089.3409758}
\bibfield{author}{\bibinfo{person}{Valerio Terragni}, \bibinfo{person}{Gunel Jahangirova}, \bibinfo{person}{Paolo Tonella}, {and} \bibinfo{person}{Mauro Pezz\`{e}}.} \bibinfo{year}{2020}\natexlab{}.
\newblock \showarticletitle{Evolutionary improvement of assertion oracles} \emph{(\bibinfo{series}{ESEC/FSE 2020})}. \bibinfo{publisher}{Association for Computing Machinery}, \bibinfo{address}{New York, NY, USA}, \bibinfo{pages}{1178–1189}.
\newblock
\showISBNx{9781450370431}
\urldef\tempurl%
\url{https://doi.org/10.1145/3368089.3409758}
\showDOI{\tempurl}


\bibitem[Tufano et~al\mbox{.}(2020)]%
        {AthenaTest}
\bibfield{author}{\bibinfo{person}{Michele Tufano}, \bibinfo{person}{Dawn Drain}, \bibinfo{person}{Alexey Svyatkovskiy}, \bibinfo{person}{Shao Deng}, {and} \bibinfo{person}{Neel Sundaresan}.} \bibinfo{year}{2020}\natexlab{}.
\newblock \bibinfo{title}{Unit Test Case Generation with Transformers}.
\newblock
\newblock
\urldef\tempurl%
\url{https://doi.org/10.48550/arXiv.2009.05617}
\showDOI{\tempurl}


\bibitem[Wang et~al\mbox{.}(2024)]%
        {10.1145/3650105.3652293}
\bibfield{author}{\bibinfo{person}{Hailong Wang}, \bibinfo{person}{Tongtong Xu}, {and} \bibinfo{person}{Bei Wang}.} \bibinfo{year}{2024}\natexlab{}.
\newblock \showarticletitle{Deep Multiple Assertions Generation}. In \bibinfo{booktitle}{\emph{Proceedings of the 2024 IEEE/ACM First International Conference on AI Foundation Models and Software Engineering}} (Lisbon, Portugal) \emph{(\bibinfo{series}{FORGE '24})}. \bibinfo{publisher}{Association for Computing Machinery}, \bibinfo{address}{New York, NY, USA}, \bibinfo{pages}{1–11}.
\newblock
\showISBNx{9798400706097}
\urldef\tempurl%
\url{https://doi.org/10.1145/3650105.3652293}
\showDOI{\tempurl}


\bibitem[Wang et~al\mbox{.}(2020)]%
        {10.1145/3386252}
\bibfield{author}{\bibinfo{person}{Yaqing Wang}, \bibinfo{person}{Quanming Yao}, \bibinfo{person}{James~T. Kwok}, {and} \bibinfo{person}{Lionel~M. Ni}.} \bibinfo{year}{2020}\natexlab{}.
\newblock \showarticletitle{Generalizing from a Few Examples: A Survey on Few-shot Learning}.
\newblock \bibinfo{journal}{\emph{ACM Comput. Surv.}} \bibinfo{volume}{53}, \bibinfo{number}{3}, Article \bibinfo{articleno}{63} (\bibinfo{date}{June} \bibinfo{year}{2020}), \bibinfo{numpages}{34}~pages.
\newblock
\showISSN{0360-0300}
\urldef\tempurl%
\url{https://doi.org/10.1145/3386252}
\showDOI{\tempurl}


\bibitem[Watson et~al\mbox{.}(2020a)]%
        {ATLAS_2020}
\bibfield{author}{\bibinfo{person}{Cody Watson}, \bibinfo{person}{Michele Tufano}, \bibinfo{person}{Kevin Moran}, \bibinfo{person}{Gabriele Bavota}, {and} \bibinfo{person}{Denys Poshyvanyk}.} \bibinfo{year}{2020}\natexlab{a}.
\newblock \showarticletitle{On learning meaningful assert statements for unit test cases}. In \bibinfo{booktitle}{\emph{Proceedings of the ACM/IEEE 42nd International Conference on Software Engineering}} \emph{(\bibinfo{series}{ICSE ’20})}. \bibinfo{publisher}{ACM}.
\newblock
\urldef\tempurl%
\url{https://doi.org/10.1145/3377811.3380429}
\showDOI{\tempurl}


\bibitem[Watson et~al\mbox{.}(2020b)]%
        {ATLAS}
\bibfield{author}{\bibinfo{person}{Cody Watson}, \bibinfo{person}{Michele Tufano}, \bibinfo{person}{Kevin Moran}, \bibinfo{person}{Gabriele Bavota}, {and} \bibinfo{person}{Denys Poshyvanyk}.} \bibinfo{year}{2020}\natexlab{b}.
\newblock \showarticletitle{On Learning Meaningful Assert Statements for Unit Test Cases}. In \bibinfo{booktitle}{\emph{2020 IEEE/ACM 42nd International Conference on Software Engineering (ICSE)}}. \bibinfo{pages}{1398--1409}.
\newblock


\bibitem[Wei et~al\mbox{.}(2023)]%
        {wei2023chainofthought}
\bibfield{author}{\bibinfo{person}{Jason Wei}, \bibinfo{person}{Xuezhi Wang}, \bibinfo{person}{Dale Schuurmans}, \bibinfo{person}{Maarten Bosma}, \bibinfo{person}{Brian Ichter}, \bibinfo{person}{Fei Xia}, \bibinfo{person}{Ed Chi}, \bibinfo{person}{Quoc Le}, {and} \bibinfo{person}{Denny Zhou}.} \bibinfo{year}{2023}\natexlab{}.
\newblock \bibinfo{title}{Chain-of-Thought Prompting Elicits Reasoning in Large Language Models}.
\newblock
\newblock
\showeprint[arxiv]{2201.11903}~[cs.CL]


\bibitem[White et~al\mbox{.}(2023)]%
        {white2023promptpatterncatalogenhance}
\bibfield{author}{\bibinfo{person}{Jules White}, \bibinfo{person}{Quchen Fu}, \bibinfo{person}{Sam Hays}, \bibinfo{person}{Michael Sandborn}, \bibinfo{person}{Carlos Olea}, \bibinfo{person}{Henry Gilbert}, \bibinfo{person}{Ashraf Elnashar}, \bibinfo{person}{Jesse Spencer-Smith}, {and} \bibinfo{person}{Douglas Schmidt}.} \bibinfo{year}{2023}\natexlab{}.
\newblock \bibinfo{title}{A Prompt Pattern Catalog to Enhance Prompt Engineering with ChatGPT}.
\newblock
\newblock
\urldef\tempurl%
\url{https://doi.org/10.48550/arXiv.2302.11382}
\showDOI{\tempurl}


\bibitem[Ye et~al\mbox{.}(2023)]%
        {ye2023satlm}
\bibfield{author}{\bibinfo{person}{Xi Ye}, \bibinfo{person}{Qiaochu Chen}, \bibinfo{person}{Isil Dillig}, {and} \bibinfo{person}{Greg Durrett}.} \bibinfo{year}{2023}\natexlab{}.
\newblock \bibinfo{title}{SatLM: Satisfiability-Aided Language Models Using Declarative Prompting}.
\newblock
\newblock
\showeprint[arxiv]{2305.09656}~[cs.CL]


\bibitem[Yu et~al\mbox{.}(2022)]%
        {yu2019automated}
\bibfield{author}{\bibinfo{person}{Hao Yu}, \bibinfo{person}{Yiling Lou}, \bibinfo{person}{Ke Sun}, \bibinfo{person}{Dezhi Ran}, \bibinfo{person}{Tao Xie}, \bibinfo{person}{Dan Hao}, \bibinfo{person}{Ying Li}, \bibinfo{person}{Ge Li}, {and} \bibinfo{person}{Qianxiang Wang}.} \bibinfo{year}{2022}\natexlab{}.
\newblock \showarticletitle{Automated Assertion Generation via Information Retrieval and Its Integration with Deep Learning}. In \bibinfo{booktitle}{\emph{Proceedings of the 44th International Conference on Software Engineering}} (Pittsburgh, Pennsylvania) \emph{(\bibinfo{series}{ICSE '22})}. \bibinfo{publisher}{Association for Computing Machinery}, \bibinfo{address}{New York, NY, USA}, \bibinfo{pages}{163–174}.
\newblock
\showISBNx{9781450392211}
\urldef\tempurl%
\url{https://doi.org/10.1145/3510003.3510149}
\showDOI{\tempurl}


\bibitem[Zamprogno et~al\mbox{.}(2023)]%
        {Autoassert}
\bibfield{author}{\bibinfo{person}{Lucas Zamprogno}, \bibinfo{person}{Braxton Hall}, \bibinfo{person}{Reid Holmes}, {and} \bibinfo{person}{Joanne~M. Atlee}.} \bibinfo{year}{2023}\natexlab{}.
\newblock \showarticletitle{Dynamic Human-in-the-Loop Assertion Generation}.
\newblock \bibinfo{journal}{\emph{IEEE Transactions on Software Engineering}} \bibinfo{volume}{49}, \bibinfo{number}{4} (\bibinfo{year}{2023}), \bibinfo{pages}{2337--2351}.
\newblock
\urldef\tempurl%
\url{https://doi.org/10.1109/TSE.2022.3217544}
\showDOI{\tempurl}


\end{thebibliography}

\end{document}